\documentclass[12pt]{article}

\usepackage{latexsym,amsmath,amssymb,theorem,epsfig,graphicx,hyperref}

\usepackage{graphicx}
\usepackage{graphicx,subfigure}
\usepackage{epstopdf}
\usepackage{amssymb}
\usepackage{amsmath}
\usepackage{psfrag}
\usepackage{epsfig}
\usepackage{cancel}
 \allowdisplaybreaks[4]

\usepackage[all]{xy}

\DeclareFontFamily{OT1}{rsfs10}{}
\DeclareFontShape{OT1}{rsfs10}{m}{n}{ <-> rsfs10 }{}
\DeclareMathAlphabet{\mathscript}{OT1}{rsfs10}{m}{n}

\numberwithin{equation}{section}

\newcommand{\ns}{\normalsize}

\newcommand{\ct}{c_s^2}

\newcommand{\kk}{{\bf k}}

\newcommand{\comb}{(1-\epsilon-\epsilon_s)}

\newcommand{\es}{\epsilon_s}
\newcommand{\eps}{\epsilon_s}
\newcommand{\ep}{\epsilon}
\newcommand{\cbt}{\bar{c}_s^2}

\newcommand{\cb}{\bar{c}_s}
\newcommand{\fnl}{f_{\rm NL}}

\def\gsim{ \lower .75ex \hbox{$\sim$} \llap{\raise .27ex \hbox{$>$}} }
\def\lsim{ \lower .75ex \hbox{$\sim$} \llap{\raise .27ex \hbox{$<$}} }
\def\be{\begin{equation}}
\def\ee{\end{equation}}
\def\bea{\begin{eqnarray}}
\def\eea{\end{eqnarray}}

\usepackage{graphicx}

\newcommand{\ba}{\begin{array}}
\newcommand{\ea}{\end{array}}

\newcommand{\cA}{{\cal A}}
\newcommand{\commentout}[1]{}
\begin{document}
\begin{titlepage}
\title{
  \hfill{\ns }  \\[1em]
   {\LARGE Non-Gaussianity in single field models \\ without slow-roll}
\\[1em] }
\author{Johannes Noller and Jo\~{a}o Magueijo
     \\[0.5em]
     {\ns Theoretical Physics, Blackett Laboratory, Imperial College, London, SW7 2BZ, UK}\\}

\date{}

\maketitle

\begin{abstract}
We investigate non-Gaussianity in general single field models without assuming slow-roll conditions or the exact scale-invariance of the scalar power spectrum. The models considered include general single field inflation (e.g. DBI and canonical inflation) as well as bimetric models. We compute the full non-Gaussian amplitude $\cA$, its size $\fnl$, its shape, and the running with scale $n_{NG}$. In doing so we show that observational constraints allow significant violations of slow roll conditions and we derive explicit bounds on slow-roll parameters for fast-roll single field scenarios. A variety of new observational signatures is found for models respecting these bounds. We also explicitly construct concrete model implementations giving rise to this new phenomenology.
\end{abstract}

\thispagestyle{empty}

\end{titlepage}

\setcounter{tocdepth}{2}
\tableofcontents

\section{Introduction}

Non-Gaussianity as probed by upcoming experiments such as Planck will be a very powerful tool for selecting models of primordial structure formation~\cite{wmapcosmo,Komatsu:2009kd,Koyama:2010xj,Komatsu:2003iq,Bartolo:2004if,Komatsu:2010hc}. One of the most interesting questions one may hope to answer is: Are single field models of structure formation sufficient, or do we need multiple fields to meet observational constraints? The hallmark of single field models that generate large non-Gaussianity is often thought to be an equilateral type non-Gaussian shape~\cite{chks,kp,chenreview,largeng,bimetricng,Li:2008qc}. Known modifications to this shape can be achieved by introducing, for example, non-Bunch-Davies vacua~\cite{chks,photo,nonbdmartin,nonbddanielsson,Ashoorioon:2010xg} or resonant running~\cite{largeng,resonant1,resonant2,foldedchen,Flauger:2010ja}. However, models with additional fields, such as multifield inflation~\cite{Liddle:1998jc,Bernardeau:2002jy,Vernizzi:2006ve,Wands:2007bd,Langlois:2008wt,Langlois:2008qf,Misra:2008tx,Rodriguez:2008hy,Byrnes:2009qy,Hotchkiss:2009pj,Byrnes:2010em,Peterson:2010mv,rpintermediate,Seery:2005gb,Mulryne:2010rp} or models adding curvaton fields~\cite{lindecurvaton,Lythcurvaton,Malik:2006pm,huangcurvaton,licurvaton,chamberscurvaton,Mazumdar:2010sa,Demozzi:2010aj,Fonseca:2011iz}, can give rise to a much greater variety of non-Gaussian signatures (see e.g.~\cite{rpintermediate}). It is therefore interesting to investigate whether multifield models are really unique in predicting all such alternative shapes.

In this paper we demonstrate that non-slow-roll single field models can produce a much richer variety of detectable non-Gaussian signatures than those following from slow-roll conditions. Typically slow-roll conditions are expected to be imposed by observational constraints derived from primordial structure formation~\cite{wmapcosmo} and from bounds arising by requiring solutions of e.g. flatness, homogeneity and isotropy problems. In~\cite{kp} it was shown that for exactly scale-invariant non-canonical models, a much wider range of slow-roll parameters is allowed than those satisfying slow-roll conditions, whilst still satisfying all observational bounds. We extend this work by dropping the requirement of exact scale-invariance, explicitly computing the corresponding new bounds for slow-roll parameters and showing that an even wider range of parameters becomes observationally viable.

Non-Gaussianities for non-slow-roll single field models have been previously computed for the case of exact scale invariance $n_s = 1$~\cite{kp} and that of an effective DBI type action~\cite{bimetricng}. 
We will generalize this work, restricting ourselves to expanding solutions as provided, for example, by inflationary~\cite{infl} and bimetric models~\cite{joao2,bimetricng}. Contracting models have been considered in the context of ekpyrotic/cyclic models~\cite{ekp,multiekp} and effective perfect fluid models~\cite{chyd,nahyd}. 
\footnote{While this paper was being finished, references~\cite{khouryrapid,baumann} appeared, which further investigate contracting models. \cite{baumann} also considers strong coupling constraints for expanding models.}
In this paper we hope to complete the picture in the expanding case by deriving non-Gaussianities for general single field models without assuming smallness of slow-roll parameters $\ep_i$ or exact scale invariance ($n_s = 1$). We investigate the phenomenology associated with such models, showing that they have several distinct observable signatures. 
These include a generically large running of non-Gaussianity $n_{NG}$ with scale, the suppression of the size of non-Gaussianity $\fnl$ by non-slow-roll $\ep_i$ and the emergence of qualitatively new shapes for single field models e.g. ones with dominant contributions in the folded limit $2k_1 = 2k_2 = k_3$.
These considerations widen the number of potential future observations that could be explained by single field models. And even if future measurements turn out to favor multifield models, it is still important to fully understand the potential effects of individual fields that may participate in such models.

The plan of the paper is as follows. In section~\ref{singlefield} we set the stage by considering the two-point correlation functions for scalar and tensor perturbations in general single field models. We compute the resulting observational constraints on slow-roll parameters and also summarize bounds coming from other sources. 
Then, in section~\ref{nongaussianity}, we compute non-Gaussianities,
explicitly giving the full amplitudes for general single field models without assuming slow-roll conditions or exact scale-invariance (section~\ref{fullamp}). Observational signatures are considered in detail for $\fnl$  (section~\ref{fnlsection}), non-Gaussianity shapes (section~\ref{shapes}) and running with scale $n_{NG}$ (section~\ref{running}). In section~\ref{concrete} concrete models are provided for the phenomenologically interesting cases by explicitly constructing their actions. In section~\ref{conc} we conclude, summarize results and point out intriguing possibilities for further work.

\section{Single field models without slow-roll} \label{singlefield}

The models we shall consider have an action of the form~\cite{kessence,garrigamukhanov}
\begin{equation} \label{general}
S=\int {\rm d}^4x \sqrt{-g} \left[\frac{R}{2} +
P(X,\phi)\right]~, 
\end{equation}
where the pressure $P$ is a general function of a single scalar field $\phi$ and its kinetic term $X=-\frac{1}{2}g^{\mu\nu}\partial_{\mu}\phi \partial_{\nu}\phi$. For a canonical field $P(X,\phi) = X - V(\phi)$. In principle $P$ can contain higher derivatives of the field $\phi$ as well, however such terms are expected to be suppressed by the UV cut-off scale in an effective field theory. This motivates studying an action of the form \eqref{general}, which is the most general Lorentz invariant action for a single scalar field minimally coupled to gravity that contains at most first derivatives of the field. The pressure $p$ and energy density $\rho$ for these models are given by
\begin{eqnarray}
p &=& P(X,\phi)\,,\nonumber \\
\rho &=& 2 X P_{,X} - P(X,\phi)~,
\end{eqnarray}
where $P_{,X} \equiv \partial P / \partial X$.

We will find it useful to characterize the non-canonical nature of \eqref{general} with the following quantities~\cite{seery}
\begin{eqnarray}
c_s^2 &=& \frac{P_{,X}}{\rho_{,X}}= \frac{P_{,X}}{P_{,X}+2X P_{,XX}}\,,\\
\Sigma&=&X P_{,X}+2X^2P_{,XX}  = \frac{H^2\epsilon}{c_s^2} ~,\\
\lambda&=& X^2P_{,XX}+\frac{2}{3}X^3P_{,XXX} ~,
\label{lambda}
\end{eqnarray}
where $c_s$ is the adiabatic speed of sound. An infinite hierarchy of slow-roll parameters $\ep$ and $\eps$ and their derivatives can then be constructed for a theory \eqref{general} with an FRW metric with scale factor $a(t)$ and corresponding Hubble rate $H(t) = {\dot a}/a$. These hierarchies are
\begin{eqnarray}
\epsilon \equiv - \frac{\dot H}{H^2},\;
\eta \equiv \frac{\dot \epsilon}{\epsilon H},... \qquad ; \qquad
\es \equiv \frac{\dot c_s}{c_s H},\;
\eta_s \equiv \frac{\dot \es}{\es H},... \label{slowroll}
\end{eqnarray}
The slow-roll approximation requires that $\ep,\eta ... \eps, \eta_s... \ll 1$ and accelerated expansion takes place when $\ep < 1$. Note that these definitions are more general than alternative definitions in terms of the scalar potential and its derivatives, which assume canonical kinetic terms. Perhaps more fittingly \eqref{slowroll} have therefore also been dubbed ``slow-variation parameters''~\cite{kinneyflow,beanconstraints,Peiris:2007gz}. DBI inflation~\cite{Alishahiha:2004eh}, for example, is slow varying in this sense yet allows steep potentials and hence the inflaton is not slow-rolling. However, in order to agree with convention and avoid confusion, we will refer to \eqref{slowroll} as slow-roll parameters from now on.

Truncating slow-roll parameters at some derivative order, setting the remainder to zero, allows us to approximate $H$ and $c_s$ with Taylor expansions.  Following~\cite{kp,bimetricng} we consider models where slow-roll conditions are broken by $\ep,\eps \sim {\cal O}(1)$. We will assume that $\ep,\eps$ are approximately constant, so that slow-roll conditions still hold for the remaining negligible parameters $\eta,\eta_s...\approx 0$.

\subsection{Observational constraints on (non)-slow-roll}

Large classes of both inflationary and bimetric models can be described by action~\eqref{general}. Bimetric models generically have $\ep > 1$ and are discussed in detail in section~\ref{bimetric}. For the inflationary branch $\ep < 1$, however, observations can be used to constrain the allowed values of slow-roll parameters $\ep,\eps$. These constraints are derived throughout the paper and we summarize the resulting bounds upfront here. The observational bounds considered are the following:

\begin{itemize}
\item {\bf The spectral index $n_s$:} The spectral index of scalar perturbations $n_s$ is computed in section~\ref{spectral}. In terms of the slow-roll parameters it can be written as
\begin{equation}
n_s - 1 = \frac{2 \epsilon + \es}{\es + \epsilon -1}\;.
\end{equation}
The WMAP bounds on $n_s$ in the presence of tensor perturbations are $n_s = 0.973 \pm 0.028$ at $2\sigma$. Note that these constraints are dependent on e.g. the precise value of $H_0$~\cite{wmapcosmo} and the physics of reionization~\cite{pandolfi}. Contrary to the canonical field case, for a general single field model~\eqref{general} the smallness of $n_s - 1 \ll 1$ no longer automatically translates into a constraint on $\ep$ alone. Instead, for any parameter $\ep$, one can find a corresponding speed of sound profile parameterized by $\eps$, which returns any desired spectral index. Bounds on $n_s$ therefore only translate into constraints on the relation between $\ep$ and $\eps$, not on either variable by itself.

\item {\bf The tensor-to-scalar ratio $r$:} The tensor-to-scalar ratio $r$ is similarly constrained by CMB observations. The WMAP experiment provides a bound $r < 0.24$ at $2\sigma$~\cite{wmapcosmo}. In section~\ref{tensor} we discuss how the first order relation $r = 16 \ep c_s$ is modified in general single field models without slow-roll and compute corresponding bounds on slow-roll parameter $\ep$. We find
\be
\ep \lesssim 0.4.
\ee 

\item {\bf The running $n_{NG}$:} Introducing non-canonical kinetic terms can lead to a strong running of non-Gaussianities with scale $n_{NG}$~\cite{kp,runningchen,runningconstraints,Byrnes:2010ft}, irrespective of whether the 2-point correlation function as measured by $n_s$ is near-scale-invariant or not. This can impose further constraints on $\ep$. If non-Gaussianities show a blue tilt, this means they peak on smaller scales. For a perturbative treatment to be applicable, one therefore needs to ensure that non-Gaussianities on the smallest observable scales are not too large for a perturbative approach to break down. In section~\ref{running} we identify under which conditions non-Gaussianities are blue tilted and what constraints on $\ep$ follow. We show that constraints are in principle model-dependent. However, for a very large subclass of models one can obtain rough bounds
\begin{eqnarray}
\ep &\lesssim& 0.3  \quad\quad \text{if} \quad\quad \fnl(\text{CMB}) \sim {\cal O}(100) \\
\ep &\lesssim& 0.5  \quad\quad \text{if} \quad\quad \fnl(\text{CMB}) \sim {\cal O}(1).
\end{eqnarray}
This result agrees with bounds obtained in the exactly scale invariant case~\cite{kp}. However, we show that these limits are in fact relatively insensitive to the precise value taken by $n_s$, but strongly depend on parameters $c_s,\Sigma,\lambda$.

\item {\bf The Big Bang problems:} Observational bounds only place constraints on primordial perturbations for a small range of scales: CMB ($\sim 10^3 {\rm Mpc}$) to galactic scales ($\sim 1 {\rm Mpc}$). $n_s$,$r$ and $n_{NG}$ as discussed above are therefore only observationally constrained over these scales. 

In an inflationary setup this corresponds to approximately 10 e-folds, where the number of e-folds $N$ of inflation is given by $d N = -H dt$. In order to resolve the flatness, homogeneity and isotropy problems of cosmology, a greater number of e-folds is needed. Conventionally this is chosen $N \sim 60$, although the minimal amount of e-folds needed is not very well defined and depends e.g. on the energy scale of inflation, details of the reheating mechanism, etc.~\cite{Eastherduration} The inflationary condition $\ep < 1$ is sufficient to yield an attractor solution towards flatness, homogeneity and isotropy~\cite{kp}. A large number of e-folds requires $\ep < 1$ for several Hubble times, which is controlled by parameter $\eta$. As we consider cases for which $\eta \sim 0 \ll 1$, the fractional change in $\ep$ per Hubble time is small and hence starting with $\ep < 1$ guarantees that a prolonged phase of inflation takes place.

However, we should note that several other solutions are possible as well. For example, the scaling phase responsible for near-scale-invariant perturbations on observable scales may be followed by a different inflationary phase (scaling or not)~\cite{kp}. Particle physics may suggest that inflation is actually a multiple step process, where one relatively short initial phase is responsible for observable structure on very large scales~\cite{stepinflation}. Inflation then ends and is followed by (an)other phase(s) of accelerated expansion later. It could also be that an altogether different mechanism\footnote{And of course, as the flatness, homogeneity and isotropy problems are essentially initial value problems, the correct initial values may be set by pre-inflationary physics. Whilst such a solution may be very non-definitive and as such less appealing, one should keep in mind that there is certainly no guarantee that all cosmological puzzles can be answered at energy scales we have (in-)direct access to.} resolves the problems in question here~\cite{vsl0,vsl1,vslreview}. 
In any case no competitive constraints on $\ep$ can be derived from consideration of these cosmological problems. Also note that bimetric models resolve these problems in a radically different way~\cite{vsl0,vsl1,vslreview}.
\end{itemize}

Inflationary models of the type envisaged by~\eqref{general}, with $\ep \lesssim 0.3$, meet all the observational bounds.

\subsection{The spectral index $n_s$} \label{spectral}

In this section we sketch the calculation of the primordial power spectrum of scalar perturbations. Scalar perturbations in the metric
\be
ds^2 = (1 + 2\Phi)dt^2 - (1 - 2\Phi)a^2(t)\gamma_{ij}dx^i dx^j,
\ee
can be analyzed in terms of the the gauge invariant curvature perturbation $\zeta$
\be
\zeta = \frac{5\rho + 3p}{3(\rho + p)}\Phi + \frac{2 \rho}{3(\rho + p)}\frac{\dot{\Phi}}{H},
\ee
where $p$ and $\rho$ are the background pressure and energy density respectively.
Perturbing \eqref{general} to quadratic order in $\zeta$ one obtains~\cite{garrigamukhanov}
\be \label{actionzeta}
S_2 = \frac{M_{\rm Pl}^2}{2} \int {\rm d}^3x{\rm d}\tau \;z^2\left[ \left(\frac{d\zeta}{d\tau}\right)^2 - c_s^2(\vec{\nabla}\zeta)^2\right]\,,
\ee
where $\tau$ is conformal time, and $z$ is defined as $z = a \sqrt{2 \epsilon}/c_s$. 

As a consequence of considering non-canonical kinetic terms, we have introduced a dynamically varying speed of sound. This means it is convenient to work in terms of the ``sound-horizon'' time ${\rm d}y = c_s {\rm d}\tau$ instead of $\tau$~\cite{kp}. For the case relevant here, i.e. $\eta = \eta_s = 0$, one finds
\be \label{y} 
y = \frac{c_s}{(\epsilon+\epsilon_s -1)aH}\,.
\ee 
In terms of $y$-time the quadratic action takes on the familiar form
\be
S_2 = \frac{M_{\rm Pl}^2}{2}\int {\rm d}^3x{\rm d}y \;q^2\left[  \zeta'^2 -(\vec{\nabla}\zeta)^2\right] \qquad ,\qquad q \equiv \sqrt{c_s}z = \frac{a \sqrt{2 \epsilon}}{\sqrt{c_s}}\,,
\label{qdef}
\ee
where $'\equiv {\rm d}/{\rm d}y$. It is useful to list the behavior in proper time $\tau$ and in $y$-time of some relevant quantities for later reference: 
\begin{eqnarray} \label{twotimes}
a \sim (-\tau)^{\frac{1}{\epsilon - 1}}\;&;\qquad c_s \sim (-\tau)^{\frac{\epsilon_s}{\epsilon - 1}}\;&; \qquad H \sim (-\tau)^{\frac{-\epsilon}{\epsilon - 1}}, \nonumber \\
a \sim (-y)^{\frac{1}{\epsilon_s +\epsilon - 1}}\;&;\qquad c_s \sim (-y)^{\frac{\epsilon_s}{\epsilon_s + \epsilon - 1}}\;&; \qquad H \sim (-y)^{\frac{-\epsilon}{\epsilon_s + \epsilon - 1}}.
\label{const} 
\end{eqnarray}

Upon quantization the perturbations are expressed through creation and annihilation operators as follows,
\begin{equation} \label{modes}
\zeta(y, \kk) = u_k(y)a(\kk) + u_k^*(y) a^\dagger(-\kk) \qquad , \qquad [a(\kk), a^\dagger(\kk')] = (2 \pi)^3 \delta^3(\kk - \kk').
\end{equation}
In terms of the canonically-normalized scalar variable $v = M_{\rm Pl} q\zeta$ the equations of motion for the Fourier modes are
\be
v''_k +\left(k^2 - \frac{q''}{q}\right)v_k = 0\,.
\label{veqn}
\ee
If $\ep,\eta,\eps,\eta_s$ are constant, as is the case here, one can solve for ${q''}/{q}$ exactly~\cite{stewart}:
\begin{equation}
\frac{q''}{q} = \frac{1}{y^2}\left(\nu^2 - \frac{1}{4}\right).
\end{equation}
The precise solution for $v_k(y)$ depends on which initial vacuum is chosen. Typically this is the Bunch-Davis vacuum, which is defined to be the state annihilated by $a(\kk)$ as $t \to - \infty$. With this vacuum choice the solution for $v_k(y)$ is
\begin{equation}
v_k(y) = \frac{\sqrt{\pi}}{2} \sqrt{- y} \, H^{(1)}_\nu (- k y).
\end{equation}
where $H^{(1)}_\nu$ are Hankel functions of the first kind. However, we note that different adiabatic vacuum choices are possible in principle and also have distinctive observational signatures (particularly non-Gaussianities)~\cite{chenreview,nonbdmartin,nonbddanielsson,chks}. For the modes defined in \eqref{modes} one finds
\begin{equation} \label{propagator}
u_k(y) =  \frac{c_s^{1/2}}{a M_{\rm Pl} \sqrt{2 \epsilon}} v_k(y) =  \frac{c_s^{1/2}}{a M_{\rm Pl} 2^{3/2}} \sqrt{\frac{\pi}{\epsilon}} \sqrt{-y} H^{(1)}_\nu (- k y) .
\end{equation}
Therefore the 2-point correlation function for $\zeta$ is:
\begin{equation} \label{2point}
\langle \zeta(\textbf{k}_1)\zeta(\textbf{k}_2)\rangle = (2\pi)^5
\delta^3(\kk_1+\kk_2)  \frac{P_\zeta}{2 k_1^3}\,,
\end{equation}
where the expression for the power spectrum $P_\zeta$ is
\begin{equation}
P_\zeta \equiv \frac{1}{2\pi^2}k^3\left\vert\zeta_k\right\vert^2 = \frac{(\epsilon_s + \epsilon-1)^2 \, 2^{2 \nu - 3}}{2 (2 \pi)^2\epsilon}\frac{{\bar H}^2}{{\bar c}_sM_{\rm Pl}^2}\,.
\label{zetaPk}
\end{equation}
Note that we have approximated the propagator $u_k(y)$ assuming $n_s - 1 \ll 1$ in accordance with observations here (for details see appendix~\ref{details} - a more general expression for $P_\zeta$ can be found in~\cite{stewart}). The bar symbol means that the corresponding quantity has to be evaluated, for each mode $k$, at sound horizon exit, i.e. when $y = k^{-1}$. The spectral index $n_s$ is finally given by
\begin{equation} \label{sindex}
n_s - 1 \equiv \frac{d \text{ln} P_\zeta}{d \text{ln} k} = 3 - 2 \nu = \frac{2 \epsilon + \es}{\es + \epsilon -1}\;.
\end{equation}
Therefore exact scale invariance obtains when $n_s = 1$, i.e. $\eps = -2 \ep$. Constraints on $n_s$ consequently translate into relational constraints on $\ep$ and $\eps$. Importantly this means no direct bound on either $\ep$ or $\eps$ can be obtained in this way, because for any parameter choice $\ep$ one can find an appropriate $\eps$ to yield a (near)scale-invariant solution. This is in contrast to the case of canonical fields, where the smallness of $n_s -1$ requires $\ep$ to be small.

\subsection{The tensor-to-scalar ratio $r$}\label{tensor}

In inflation~\cite{infl} primordial quantum tensor fluctuations are sourced on sub-horizon scales before leaving the horizon and freezing out in analogy to scalar fluctuations. Present observational upper bounds on the tensor-to-scalar ratio $r$ therefore constrain the amount of gravity waves that could have been generated in the early universe. Here we use bounds on $r$ to constrain slow-roll parameter $\ep$.
Note that bimetric models of the primordial epoch do not resolve the horizon problem for gravitational waves~\cite{joao2}. As no quantum tensor fluctuations are amplified to become classical outside the horizon one does not expect any significant primordial gravitational wave contribution for these models. Such models therefore predict $r \approx 0$.

The action governing tensor modes is~\cite{stewart}
\be
S = {1\over2}\int d^3{\bf k} \sum_{n = 1}^2 \int \left({\left| {\tilde{v}^{'}}_{{\bf k},n} \right|}^2 - \left(k^2 - \frac{a^{''}}{a} \right){\left| {\tilde{v}}_{{\bf k},n} \right|}^2\right)d\eta
\ee
where the sum runs over the two distinct polarization modes and we have introduced $\tilde{v}$ to distinguish between variables $\tilde{v}$ and ${v}$ associated with tensor and scalar perturbations respectively. The action is written in terms of a Mukhanov-Sasaki variable ${\tilde v}_{{\bf k},n} = \frac{a(\eta) M_{Pl}}{2} h_{{\bf k},n}$, where $a$ is the scale factor of the FRW metric. Upon quantization $\tilde{v}$ can be expressed in terms of creation and annihilation operators as
\be
\hat{{\tilde{v}}}_{{\bf k},n} = {\tilde v}_k(\eta) \hat{{\tilde{a}}}_{{\bf k},n} + {\tilde v}_k^{*}(\eta) \hat{{\tilde{a}}}_{{\bf k},n}^{\dagger}.
\ee
The equation of motion for tensor modes can then be written as
\be
{\tilde v}_k^{''} + \left( k^2 - \frac{a^{''}}{a} \right) {\tilde v}_k = 0.
\ee
Accordingly we can now write down the power spectrum of tensor perturbations, where the sum again runs over polarization modes 
\be
P_h = \frac{k^3}{2 \pi^2} \sum_n^2 \langle {\left| h_{{\bf k},n} \right|}^2 \rangle,
\ee
which can be found to be~\cite{beanconstraints,stewart}
\begin{eqnarray} \label{tensorpower}
{P}_{h} = 2^{2\mu-3} \left| \frac{\Gamma(\mu)}{\Gamma(3/2)} \right|^{2} (1-\epsilon)^{2\mu-1} \left| \frac{\sqrt{2}H}{\pi M_{pl}} \right|^{2}_{k = aH},
\end{eqnarray}
where 
\be
\mu = \frac{1}{1 - \ep} + \frac{1}{2}.
\ee
Importantly the tensor power spectrum is evaluated at horizon crossing for the tensor modes. For general theories with a non-canonical kinetic term~\eqref{general} this horizon is different from the corresponding horizon for scalar perturbations.
The spectral index of tensor perturbations $n_t$ then is
\be \label{nt}
n_t = \frac{-2 \ep}{1 - \ep}. 
\ee 
The tensor-to-scalar ratio $r$ is given by
\begin{eqnarray} \label{rapprox}
r \approx 2^{2 \mu - 3} \frac{(1 - \ep)^{2\mu - 1}}{(\eps + \ep - 1)^2}{\left| \frac{\Gamma(\mu)}{\Gamma\left({3\over2}\right)} \right|}^2 16 c_s(k_\zeta) \ep {\left( \frac{H_h}{H_\zeta} \right)}^2
\end{eqnarray}
where $c_s$ has to be evaluated at scalar horizon crossing and $H_h$ and $H_\zeta$ correspond to the Hubble factor at freeze-out for tensor and scalar modes respectively~\cite{beanconstraints}.\footnote{As a comparison with~\eqref{zetaPk} shows, we have assumed $n_s - 1 = 3 - 2\nu \ll 1$ here, i.e. near-scale-invariance.} Note that the dependence on the ratio of $H$ at different horizon crossings can induce a strong scale dependence for $r$.

Present-day observational constraints limit $r \lesssim 0.3$. The WMAP 95 $\%$ confidence level bound is $r < 0.24$~\cite{wmapcosmo}.\footnote{This bound is obtained assuming $n_t = \frac{-r}{8}$, however. The relation between $n_t$ and $r$ is of course more involved for general single field models as a comparison between~\eqref{nt} and~\eqref{rapprox} shows. However, given that we obtain a red tilt of tensor modes (and $P_h$ therefore peaks on the largest scales) the bound on $r$ should only weakly depend on the precise value of the tilt.} For a given value of $n_s$ this can therefore be used to constrain $\ep c_s (k_\zeta)$. Ref~\cite{lorenz} perform an MCMC sampling for general single field models~\eqref{general} to find $\ep c_s < 0.023$ at $2\sigma$ for a flat prior (note that the result is somewhat prior dependent). For a horizon crossing sound speed of $\cb = 0.1$ this corresponds to $\ep \lesssim 0.23$ which agrees with the approximate bound derived by~\cite{kp}. However, as we will show in the following section single field models with lower sound speeds $\cb \sim 0.05$ can satisfy current bounds on the size of non-Gaussianity, so that this bound on $\ep$ can be relaxed to obtain
\be
\ep \lesssim 0.4 
\ee
for general single field models as considered here~\eqref{general}.

\section{Non-Gaussianity} \label{nongaussianity}

In this section we compute the 3-point correlation function for the curvature perturbation $\zeta$. Perturbing \eqref{general} to cubic order in $\zeta$ one obtains the effective action derived in~\cite{seery,chks}. The result is valid outside of the slow-roll approximation and for any time-dependent sound speed:
\begin{eqnarray} \label{action3}
S_3&=&{M_{Pl}}^2\int {\rm d}t {\rm d}^3x \left\{
-a^3 \left[\Sigma\left(1-\frac{1}{c_s^2}\right)+2\lambda\right] \frac{\dot{\zeta}^3}{H^3}
+\frac{a^3\epsilon}{c_s^4}(\epsilon-3+3c_s^2)\zeta\dot{\zeta}^2 \right.
\nonumber \\ &+&
\frac{a\epsilon}{c_s^2}(\epsilon-2\es+1-c_s^2)\zeta(\partial\zeta)^2-
2a \frac{\epsilon}{c_s^2}\dot{\zeta}(\partial
\zeta)(\partial \chi) \nonumber \\ &+& \left.
\frac{a^3\epsilon}{2c_s^2}\frac{d}{dt}\left(\frac{\eta}{c_s^2}\right)\zeta^2\dot{\zeta}
+\frac{\epsilon}{2a}(\partial\zeta)(\partial
\chi) \partial^2 \chi +\frac{\epsilon}{4a}(\partial^2\zeta)(\partial
\chi)^2+ 2 f(\zeta)\left.\frac{\delta L}{\delta \zeta}\right\vert_1 \right\} ~,
\end{eqnarray}
where dots denote derivatives with respect to proper time $t$, $\partial$ is a spatial derivative, and $\chi$ is defined as
\begin{equation}
\partial^2 \chi = \frac{a^2 \epsilon}{c_s^2}\dot{\zeta}\,.
\end{equation}
Note that the final term in the cubic action, $2 f(\zeta) \frac{\delta L}{\delta\zeta}|_1$ is proportional to the linearized equations of motion and can be absorbed by a field redefinition $\zeta \rightarrow \zeta_n+f(\zeta_n)$ - for details see appendix~\ref{details}.

At first order in perturbation theory and in the interaction picture, the 3-point function is given by~\cite{malda,seery,chks}
\begin{equation} \label{interaction}
\langle
\zeta(t,\textbf{k}_1)\zeta(t,\textbf{k}_2)\zeta(t,\textbf{k}_3)\rangle=
-i\int_{t_0}^{t}{\rm d}t^{\prime}\langle[
\zeta(t,\textbf{k}_1)\zeta(t,\textbf{k}_2)\zeta(t,\textbf{k}_3),H_{\rm int}(t^{\prime})]\rangle ~,
\end{equation}
where $H_{\rm int}$ is the Hamiltonian evaluated at third order in the perturbations and follows from~\eqref{action3}. Vacuum expectation values are evaluated with respect to the interacting vacuum $|\Omega \rangle$. 

The 3-point correlation function is conventionally expressed through the amplitude ${\cal A}$
\begin{equation} \label{amplidef}
\langle \zeta(\textbf{k}_1)\zeta(\textbf{k}_2)\zeta(\textbf{k}_3)\rangle = (2\pi)^7
\delta^3(\kk_1+\kk_2+\kk_3) P_\zeta^{\;2} \frac{1}{\Pi_j k_j^3}{\cal A}\,.
\end{equation}
Also by convention the power spectrum $P_\zeta$ in the above formula is calculated for the mode $K = k_1 + k_2 + k_3$. By using~\eqref{modes} we can now calculate the amplitudes for each term appearing in the action~\eqref{action3}. Using a self-evident notation the overall amplitude is therefore given by
\be 
{\cal A} = {\cal A}_{\dot\zeta^3} + {\cal A}_{\zeta \dot\zeta^2} + {\cal A}_{\zeta(\partial \zeta)^2} + {\cal A}_{\dot \zeta \partial \zeta \partial \chi} + {\cal A}_{\epsilon^2},
\ee
where ${\cal A}_{\epsilon ^2}$ accounts for the $\partial \zeta \partial \chi \partial^2 \chi$ and  $(\partial^2 \zeta) (\partial \chi)^2$ terms in the action~\eqref{action3}.

In the following subsections we first present the main result (Sec.~\ref{fullamp}), which is a general expression for $\cA$ where $\ep$ is {\it not} constrained to satisfy $\ep \ll 1$ are given. We then study three physically interesting properties of these three-point functions: The size of the amplitude as parameterized by the variable $f_{NL}$ (Sec.~\ref{fnlsection}), its shape (Sec.~\ref{shapes}) and the running of the non-Gaussianity $n_{NG}$ (Sec.~\ref{running}). Current observational bounds are also discussed for these properties.

\subsection{The full amplitudes}\label{fullamp}

We first present results for the full non-Gaussian amplitude $\cA$. Details of the calculation are given in appendix~\ref{details}. We stress that our results do not assume slow-roll conditions for $\ep,\eps$ and are therefore valid even when these parameters become $\ep,\eps \sim {\cal O}(1)$. Providing the full amplitudes here will hopefully prove useful for the computation of non-slow-roll non-Gaussianities for specific single field models in the future. For the exactly scale-invariant case analogous non-slow-roll results were presented in~\cite{kp} and for an effective DBI type action these have been derived in~\cite{bimetricng} without assuming scale-invariance. However, as we will see in the following subsections, it is precisely the part of the amplitude that vanishes in DBI type models that is instrumental for much of the new phenomenology we find for general slow-roll violating single field models~\eqref{general}. 

The following variables will be useful in computing amplitudes
\be
\alpha_1  =  n_s - 1= 3 - 2\nu = \frac{2 \epsilon + \es}{\es + \epsilon -1} \qquad ; \qquad \alpha_2   =  \frac{2 \epsilon - \es}{\es + \epsilon -1} \label{alphas}
\ee
Also note that we can write the parameter $\lambda$ as~\cite{seery,kp}
\be
\lambda = \frac{\Sigma}{6}\left( \frac{2 f_X - 1}{\ct} -1 \right)
\ee
where
\be
f_X = \frac{\ep \eps}{3 \ep_X} \qquad , \qquad \ep_X = -\frac{\dot{X}}{H^2}\frac{\partial H}{\partial X},
\ee
and we note that there is in principle no dynamical requirement fixing $\ep_X$ (and hence $f_X$) to be small or large, even in the presence of constraints for $\ep$~\cite{seery}. For DBI models (discussed in~\ref{dbisection}) $f_X^{\text{DBI}} = 1 - \ct$, so consequently the first term in the action~\eqref{action3} vanishes and there is no ${\cal A}_{\dot\zeta^3}$ contribution to the amplitude. For the computation of ${\cal A}_{\dot\zeta^3}$ we otherwise assume for simplicity that $f_X$ (and hence the kinetic part of $\ep$, $\ep_X$) is constant.
Computing the amplitude in this way for each term in the action~\eqref{action3} we obtain
\begin{align} 
{\cal A}_{\dot\zeta^3}\, =&\ 
\frac{1}{2 {\bar c}_s^{2}} \left(\frac{k_1k_2k_3}{2 K^3}\right)^{n_s - 1} \left[(\epsilon + \epsilon_s - 1) (f_X - 1){\cal I}_{\dot\zeta^3}(\alpha_2) + {\bar c}_s^2 (\epsilon + \epsilon_s - 1){\cal I}_{\dot\zeta^3}(\alpha_1)   \right] \nonumber\;; \\[2.5mm] 
\nonumber
{\cal A}_{\zeta \dot\zeta^2}\, =&\ 
\frac{1}{4 {\bar c}_s^{2}} \left(\frac{k_1k_2k_3}{2 K^3}\right)^{n_s - 1} \left[(\epsilon - 3) {\cal I}_{\zeta \dot\zeta^2}(\alpha_2) + 3 {\bar c}_s^2 {\cal I}_{\zeta \dot\zeta^2}(\alpha_1)   \right] \nonumber\;; \\[2.5mm] \nonumber
{\cal A}_{\zeta(\partial \zeta)^2} =& \
\frac{1}{8 {\bar c}_s^{2}} \left(\frac{k_1k_2k_3}{2 K^3}\right)^{n_s - 1}  \left[(\epsilon - 2 \es +1) {\cal I}_{\zeta(\partial \zeta)^2}(\alpha_2) - {\bar c}_s^2 {\cal I}_{\zeta(\partial \zeta)^2}(\alpha_1)   \right] \nonumber\;; \\[2.5mm] \nonumber
{\cal A}_{\dot \zeta \partial \zeta \partial \chi} =& \ \nonumber
\frac{  1}{4 {\bar c}_s^{2}} \left(\frac{k_1k_2k_3}{2 K^3}\right)^{n_s - 1}  \left[ - \epsilon \, {\cal I}_{\dot \zeta \partial \zeta \partial \chi}(\alpha_2) \right] \;; \\[2.5mm]
{\cal A}_{\epsilon^2} =& \ \frac{1}{16 {\bar c}_s^2} \left(\frac{k_1k_2k_3}{2 K^3}\right)^{n_s - 1}  \left[\epsilon^2\, {\cal I}_{\epsilon^2 }(\alpha_2)\right]\,, \label{amplitudes1}
\end{align}
where
\begin{align} 
{\cal I}_{\dot\zeta^3}(\alpha)\, =&\ \cos\frac{\alpha \pi}{2} \Gamma(3+\alpha)\frac{k_1^2k_2^2k_3^2}{K^3} \nonumber\;; \\[2.5mm] 
\nonumber
{\cal I}_{\zeta \dot\zeta^2}(\alpha)\, =&\ \cos\frac{\alpha \pi}{2} \Gamma(1+\alpha) \left[(2+\alpha) \frac{1}{K}\sum_{i<j}k_i^2k_j^2  - (1+\alpha) \frac{1}{K^2} \sum_{i\neq j} k_i^2 k_j^3\right] \nonumber\;; \\[2.5mm] \nonumber
{\cal I}_{\zeta(\partial \zeta)^2}(\alpha) =& - \cos\frac{\alpha\pi}{2} \Gamma(1+\alpha) \left(\sum_i k_i^2\right)\left[\frac{K}{\alpha -1} + \frac{1}{K} \sum_{i<j}k_i k_j + \frac{1+\alpha}{K^2} k_1 k_2 k_3\right]\\[2.5mm]
= & \cos\frac{\alpha\pi}{2} \Gamma(1+\alpha)\left[\frac{1}{1-\alpha} \sum_j k_j^3 +\frac{4+2\alpha}{K}\sum_{i<j}k_i^2k_j^2 \nonumber
 - \frac{2+2 \alpha}{K^2} \sum_{i\neq j} k_i^2 k_j^3 
 \right.\\
 &  \nonumber \left. 
 + \frac{\alpha}{(1-\alpha)} \sum_{i\neq j} k_i k_j^2  - \alpha k_1 k_2 k_3 \right] \;;  \\[2.5mm]
{\cal I}_{\dot \zeta \partial \zeta \partial \chi}(\alpha) =& \cos\frac{\alpha \pi}{2}\Gamma(1+\alpha) \left[\sum_j k_j^3  + \frac{\alpha -1}{2}\sum_{i\neq j}k_i k_j^2 - 2 \frac{1+\alpha}{K^2}\sum_{i\neq j}k_i^2 k_j^3 - 2 \alpha k_1 k_2 k_3\right]\;; \nonumber \\[2.5mm]
{\cal I}_{\epsilon^2} (\alpha)= &  \cos\frac{\alpha \pi}{2} \Gamma(1+\alpha)(2+\alpha/2)\left[\sum_j k_j^3 - \sum_{i\neq j} k_i k_j^2 + 2  k_1 k_2 k_3\right]\,, \label{amplitudes2}
\end{align}
where we remind ourselves that ${\cal A}_{\epsilon ^2}$ accounts for the $\partial \zeta \partial \chi \partial^2 \chi$ and  $(\partial^2 \zeta) (\partial \chi)^2$ terms in the action~\eqref{action3}. $K = k_1 + k_2 + k_3$ is the sum of the lengths of the three wavevectors ${\bf k_1,k_2,k_3}$. 
Note that, in deriving the above results, the propagator $u_k(y)$ has been approximated in the $|k y| \ll 1$ limit (for details see appendix~\ref{details}, especially equation~\eqref{umode}). The expressions derived here reduce to the correct amplitudes in the DBI~\cite{bimetricng}, scale-invariant~\cite{kp} and slow-roll~\cite{chks} limits.

\subsection{$f_{NL}$} \label{fnlsection}

\begin{figure}[h] 
\begin{center}$
\begin{array}{ccc}
\includegraphics[width=0.3\linewidth]{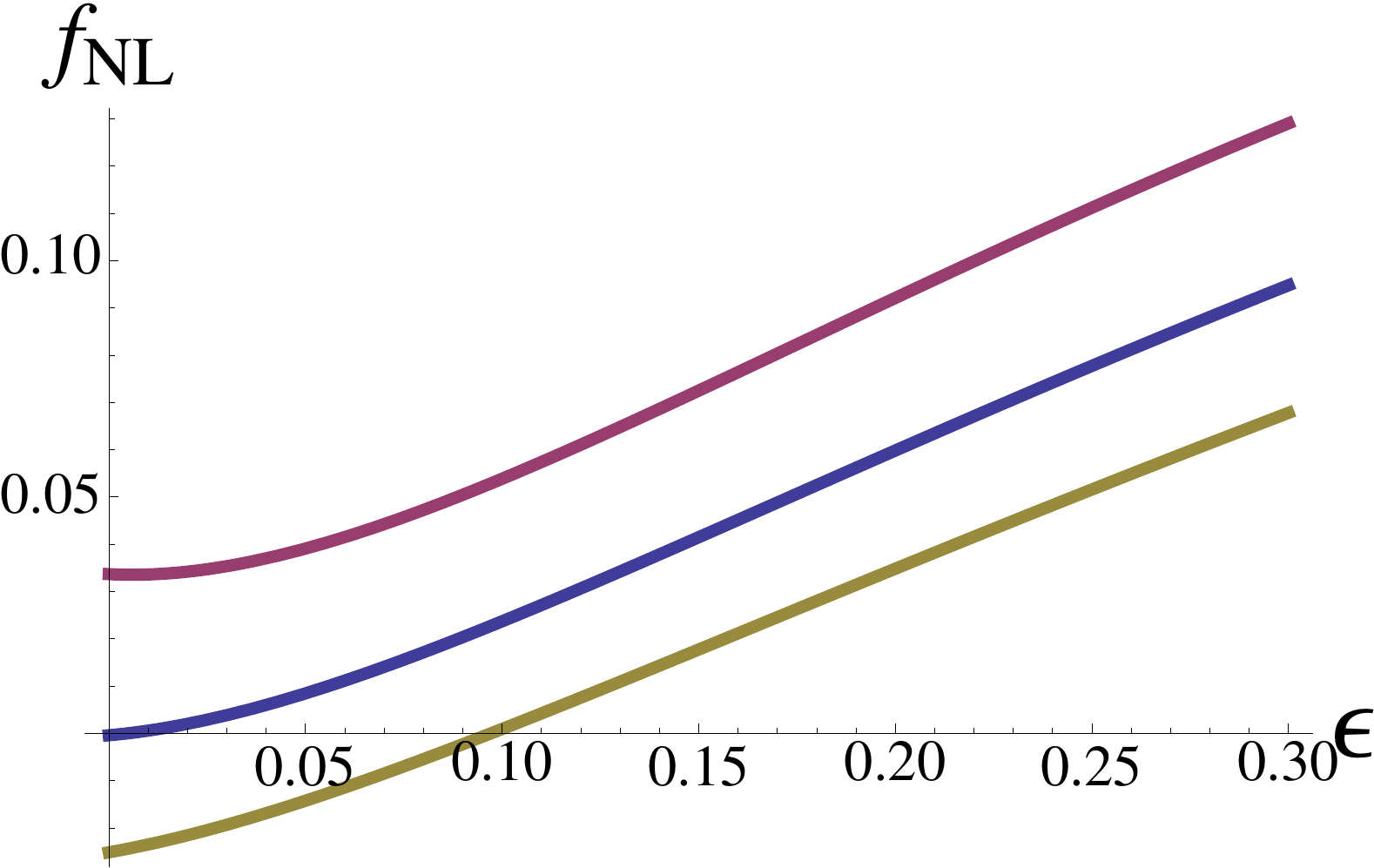} &
\includegraphics[width=0.3\linewidth]{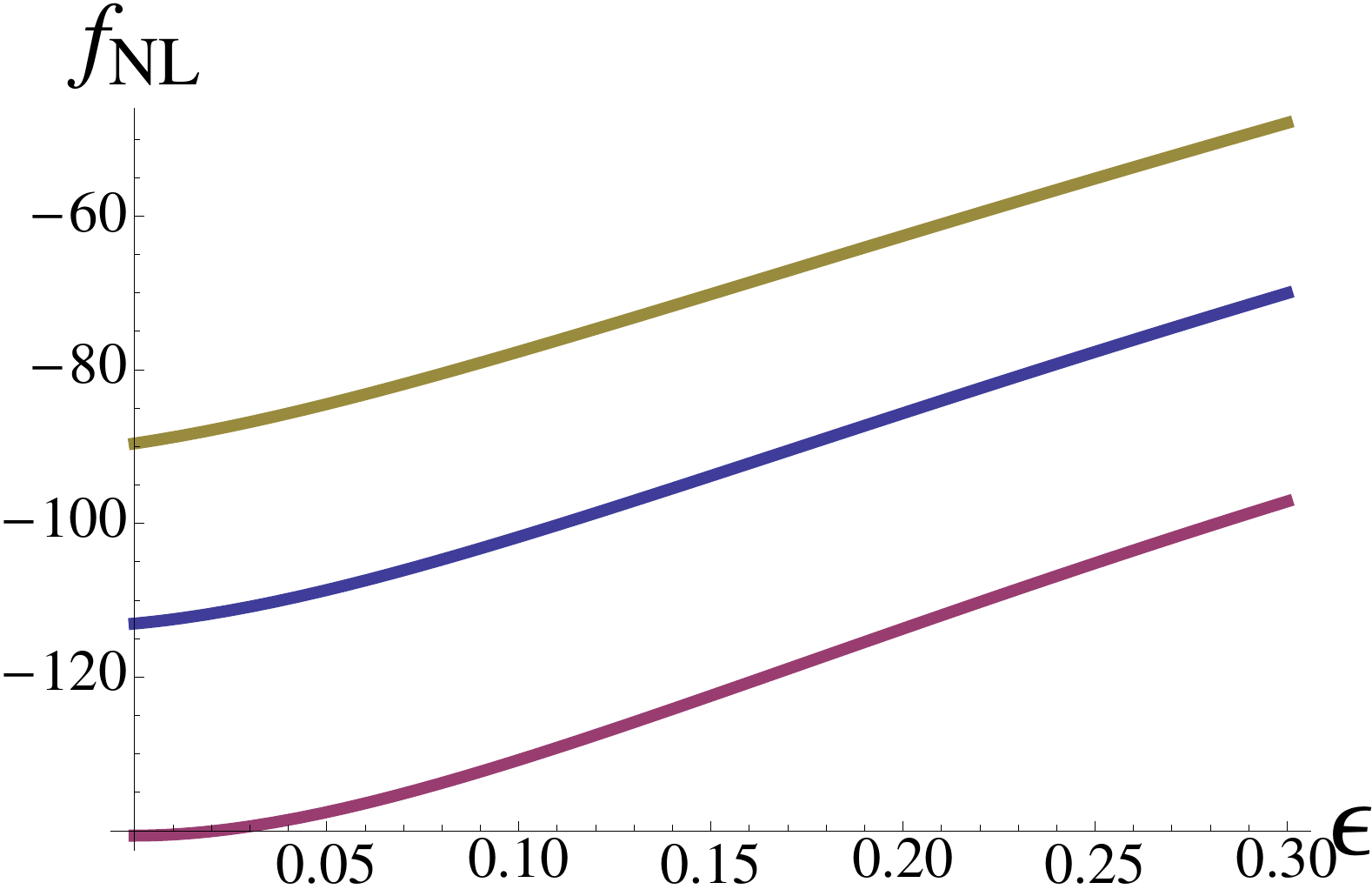} &
\includegraphics[width=0.3\linewidth]{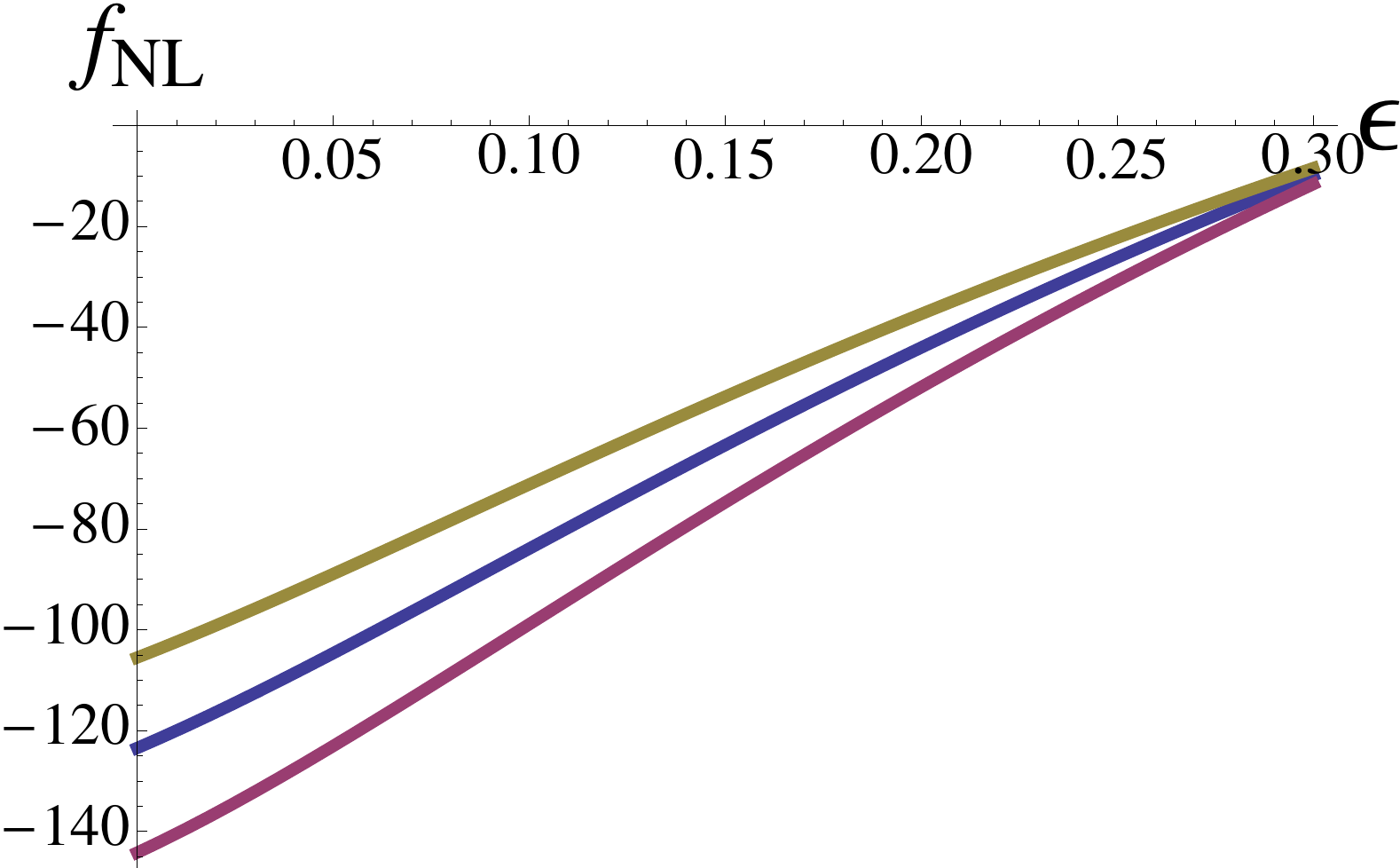}
\end{array}$
\end{center}
\caption{ $\fnl$ is plotted for parameters {\bf Left:} $\cb = 1$, $f_X = 0.01$ and $n_s = 0.96,1,1.04$ from top to bottom respectively. {\bf Middle:}  $\cb = 0.05$, $f_X = 0.01$ and $n_s = 0.96,1,1.04$ from top to bottom respectively.   {\bf Right:} $\cb = 1$, $f_X = 3 \cdot 10^3$ and $n_s = 0.96,1,1.04$ from top to bottom respectively.
} 
\label{fnlfigure}
\end{figure}

The size of the three-point correlation function is usually quoted using parameter $f_{\rm NL}$. 
In terms of the amplitudes $\cA$ this is conventionally defined as~\cite{kp}:
\be
f_{\rm NL} = 30\frac{{\cal A}_{k_1=k_2=k_3}}{K^3}\,,\label{fnl}
\ee
where we have matched amplitudes at $k_1=k_2=k_3=K/3$, i.e. in the equilateral limit. Defined in this way $\fnl$ is essentially the value of the amplitude $\cA$ at a particular point in $k$-space. Note that we follow the WMAP sign convention here, where positive $f_{\rm NL}$ physically corresponds to negative-skewness for the temperature fluctuations. In the literature expressions for $\fnl$ for a general theory with action~\eqref{general} can be found for two cases: ({\bf I}) For slowly varying $c_s$ and to first order in slow-variation parameters~\cite{chks}. ({\bf II}) For (potentially) rapidly varying $c_s$ when the 2-point correlation function is exactly scale invariant, i.e. $n_s = 1$~\cite{kp}. Here we generalize these results for an arbitrary small tilt $n_s$ and to arbitrary order in slow-variation parameter $\ep$. Hence we will focus on the novel effects on $\fnl$ due to the departure from scale-invariance and slow-roll.

An exact, but rather cumbersome, expression for $\fnl$ obtained by combining \eqref{fnl},\eqref{amplitudes1} and \eqref{amplitudes2} is given in the appendix~\ref{generalfnl}. In the limit of exact scale invariance we recover equation 8.4 of~\cite{kp} for which the following is a fitting formula which illustrates the parametric dependence of $\fnl$ in this case
\be
\fnl^{SI} = 0.27 - \frac{0.164}{\cbt} - (0.12 + 0.04 f_X)\frac{1 + \alpha_2}{\cbt}, \label{fitting}
\ee
where $\alpha_2$ is defined in~\eqref{alphas}. In order to establish a connection between different conventions we remind ourselves that $f_X$ can be expressed in terms of $\lambda/\Sigma$~\cite{seery}, given that
\be
f_X = \frac{1}{2}\left(\left(6\frac{\lambda}{\Sigma} + 1\right)\ct -1 \right). \label{fxlambda}
\ee
It follows that $|f_X| \gg 1$ in the presence of $\ct \le 1$ entails $\left|\frac{\lambda}{\Sigma}\right| \gg 1$. From \eqref{fitting} a good order of magnitude estimate for models with a large amount of non-Gaussianity as measured by $\fnl \gg 1$ is
\be
\fnl \sim {\cal O}(c_s^{-2}) + {\cal O}(\frac{\lambda}{\Sigma}). \label{oom}
\ee
Interestingly, this result corresponds to the case where fluctuations in the scalar field dominate over those in gravity~\cite{Alishahiha:2004eh,Gruzinov:2004jx}. Equation~\eqref{oom} suggests that deviations from slow-roll and breaking of scale invariance alone are not sufficient to generate significant levels of non-Gaussianity by themselves. The left graph of figure~\ref{fnlfigure} confirms this, as it shows that, in the observationally allowed range, effects from $n_s$ and $\ep$ alone can yield $\fnl \sim {\cal O}(1)$ at most.

However, departures from slow-roll and scale-invariance can have a significant effect for non-Gaussianities of detectable size $\fnl \gsim {\cal O}(1)$. Note that, considering the parametric dependence of the amplitude, $\ep_s$ is fixed in terms of $n_s,\ep$ via~\eqref{alphas}. For a given $n_s$, an $\fnl$ increasing with larger $\ep$ therefore corresponds to $\ep_s$ decreasing whilst $\fnl$ increases and vice versa. The middle and right graphs in figure~\ref{fnlfigure} show how considering non-slow-roll $\ep$ can lead to a suppression of $\fnl$. If non-Gaussianity is primarily sourced by $c_s \ll 1$ (the middle graph in~\ref{fnlfigure}), we can see that effects from non-slow-roll can suppress $\fnl$ by about a factor of 2. Breaking of exact scale invariance similarly has the observationally relevant effect of changing the size by up to $d \fnl \sim 20$.

Even more striking is the case when a large amplitude $\cA$ is primarily due to $\lambda/\Sigma \gg 1$ depicted in the right graph in figure~\ref{fnlfigure}. Then, the exact expression for $\fnl$ simplifies considerably and we obtain (irrespective of whether $c_s$ is large or small)
\be \label{fnlfx}
\fnl^{f_X \gg 1} \sim \frac{1 + \ep}{n_s - 2} \frac{f_X}{\cbt} \text{cos} \left(\alpha_2 \frac{\pi}{2}\right) \Gamma (3 + \alpha_2).
\ee
At first sight a good estimate for $\fnl$ in this case is $\fnl \sim {\cal O} (f_X / \ct) \sim {\cal O}(\frac{\lambda}{\Sigma})$ in agreement with \eqref{oom}. However, the amplitude is suppressed by a factor of $\cos\frac{\alpha_2 \pi}{2} \Gamma(3+\alpha_2)$. The suppression is so strong that the size of the amplitude can be reduced by more than an order of magnitude.
In fact, if $\ep$ is allowed to range all the way up $\ep \sim 0.4$, this effect will even lead to a change in sign for $\fnl$.
Thus slow-roll models with an $\fnl$ that violate present bounds (see below) can be reconciled with observations when non-slow-roll effects are taken into consideration.

Another interesting limit to consider is that of large speed of sound $c_s \gg 1$~\cite{vsl0,vsl1,vslreview,nahyd,chyd}. Theories with this behavior are addressed in more detail in section~\ref{bimetric}, especially the question of how to interpret an apparently ``superluminal'' speed of sound~\cite{vikman}. Concentrating on $\fnl$ for the time being, one can expand at first order in $n_s - 1$ in the limit $c_s \to \infty$ to obtain
\be \label{fnlbimetric}
\fnl^{\bar{c}_s \to \infty} \sim 0.28 - 0.04 \ep -(1.19 + 0.08 \ep) (n_s -1),
\ee
where we stress that the $f_X$-dependence is suppressed. For a spectral tilt of $n_s = 0.96$ this behaves as $\fnl \sim 0.331 - 0.038 \ep$.\footnote{Note that whilst a bound $\fnl > 5$ is often quoted~\cite{Komatsu:2001rj,Verde:1999ij}, non-Gaussianity from models with lower $\fnl$ may be detected depending on the shape, if it has a significant amplitude in regions not probed by $\fnl$. As a unique shape template is provided by models with $c_s \gg 1$, deriving constraints for the full particular shape may be a promising strategy here and allow detection of low $\fnl$ signatures.} Whilst no $\fnl \gg 1$ is obtained in accordance with~\eqref{oom}, this shows that non-Gaussianities in excess of the levels yielded by slow-roll inflation ($\fnl^{sr} \lesssim 0.01$) are possible here.

Observational constraints on $\fnl$ depend on which non-Gaussian shape is being probed~\cite{Babich:2004gb,fergusson,Komatsu:2001rj}, as discussed in the next subsection. Frequently these constraints are quoted for three shapes. Firstly, the equilateral shape, peaking in the limit $k_1 = k_2 = k_2$ and going to zero in the flat limit $k_1 + k_2 = k_3$. This roughly corresponds to the shapes following from ${\cal A}_{\dot \zeta \partial \zeta \partial \chi}, {\cal A}_{\epsilon^2}$ and also approximately ${\cal A}_{\dot\zeta^3}$ (although for ${\cal A}_{\dot\zeta^3}$ there are small differences in the flat limit $k_1 + k_2 = k_3$). A second possibility is the local shape, which peaks in the squeezed limit $k_1 = k_2 \gg k_3$ and goes to zero in the equilateral limit. This is loosely associated with the shapes of ${\cal A}_{\zeta \dot\zeta^2}$ and  ${\cal A}_{\zeta(\partial \zeta)^2}$. Finally, one may consider the orthogonal shape, which peaks in the folded configuration $2k_1 = 2k_2 = k_3$ and goes to zero in the equilateral and squeezed limits. This corresponds to the difference between ${\cal A}_{\dot\zeta^3}$ and a more exact equilateral shape such as ${\cal A}_{\dot \zeta \partial \zeta \partial \chi}$ or ${\cal A}_{\epsilon^2}$. For a list of possible factorizable shape ans\"{a}tze we refer to~\cite{Creminelli:2005hu,senatoreorthogonal,chenreview}.\footnote{A complete basis capable of describing arbitrary shapes in k-space in principle has infinitely many ``shape-members''. However, for the models considered here these three basis shapes will be sufficient.} Current WMAP limits on $\fnl$ for each of these shapes are~\cite{wmapcosmo}
\begin{eqnarray}
f_{\rm NL}^{\rm equil} &=& 26 \pm 240 \quad (95 \% \rm \; CL), \nonumber \\
f_{\rm NL}^{\rm local} &=& 32 \pm 42 \quad (95 \% \rm \; CL), \nonumber \\
f_{\rm NL}^{\rm orth}  &=& -202 \pm 208 \quad (95 \% \; \rm CL).
\end{eqnarray}

\subsection{Shapes} \label{shapes}

In the previous section the information contained in $\cA$ was condensed into a single parameter, $\fnl$. However, there is vastly more information available, encoded in the functional dependence of $\cA$~\cite{Babich:2004gb,fergusson,Komatsu:2001rj}. This is uniquely specified by the four parameters $\ep,n_s,\cb,f_X$.\footnote{Note that there is an issue of choice here as we have six variables $\ep,\eps,n_s,\cb,f_X,\lambda/\Sigma$ and two constraint equations \eqref{sindex} and \eqref{fxlambda}. For instance specifying $\ep,\eps,\cb,\lambda/\Sigma$ therefore also fixes the amplitude.} In order to disentangle effects from a possible running of $\cA$ with wavenumber $K$ and the shapes themselves, in this section we present the shape of the amplitude at fixed $K$. In interpreting the shapes shown in this section it may be useful to emphasize that from~\eqref{fnl} it follows that a good estimate for the size $\fnl$ corresponding to a given plotted shape (we normalize one wavevector $k$ and hence plot ${\cal A}(1,k_1,k_2) / (k_1 k_2)$) is the value of the plotted amplitude in the equilateral limit.

In the near future the shape and running of $\cA$ will hopefully become experimentally accessible and, if a significant amount of non-Gaussianity is detected, will greatly constrain models of structure formation in the early universe. After all, at present these models are only observationally constrained to fit or respect upper limits on a number of single value parameters (The amplitude of scalar perturbations $A$, its spectral tilt $n_S$, $\fnl$, the tensor-to-scalar ratio $r$). Fitting a full function would prove far more challenging and provide a strong model selection tool.

The prototypical shape that has been associated with significant non-Gaussianity in single-field models is one of the equilateral type~\cite{chks,kp,chenreview,largeng,bimetricng}, where the amplitude peaks in the limit $k_1 = k_2 = k_3$. This is in contrast to e.g. multifield models which can also give rise to strong local contributions to $\cA$ (i.e. peaking in the squeezed limit $k_1 \approx k_2 \gg k_3$)~\cite{chenreview} via superhorizon interactions local in position space.
This is because superhorizon evolution is local in space, as different regions are not causally connected with each other. Therefore any amount of non-Gaussianity generated after horizon exit is also local in position space and hence not local in $k$-space. It consequently peaks in the squeezed limit. In single field models, on the other hand, $\zeta$ is frozen in upon horizon exit, so no superhorizon evolution takes place. Far inside the horizon modes typically oscillate and their contributions to $\cA$ average out. For known exceptions to this see~\cite{nonbdmartin,nonbddanielsson,chks,largeng,resonant1,resonant2}. Thus $\cA$ is almost completely sourced by modes exiting the horizon at similar times and hence with similar wavelengths. Consequently $\cA$ peaks in the equilateral limit.

\begin{figure}[h] 
\begin{center}$
\begin{array}{ccc}
\includegraphics[width=0.3\linewidth]{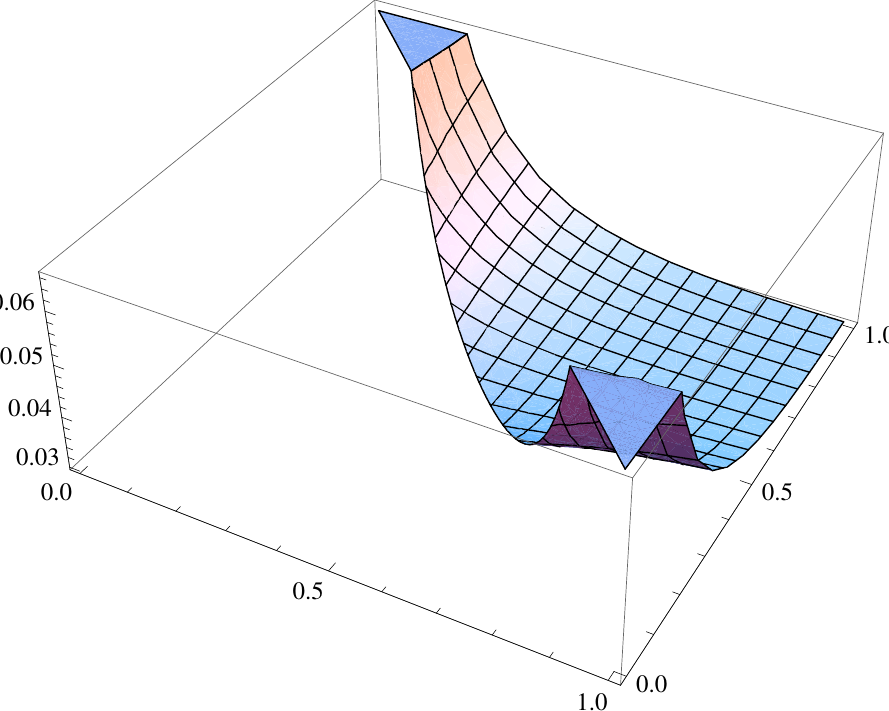} &
\includegraphics[width=0.3\linewidth]{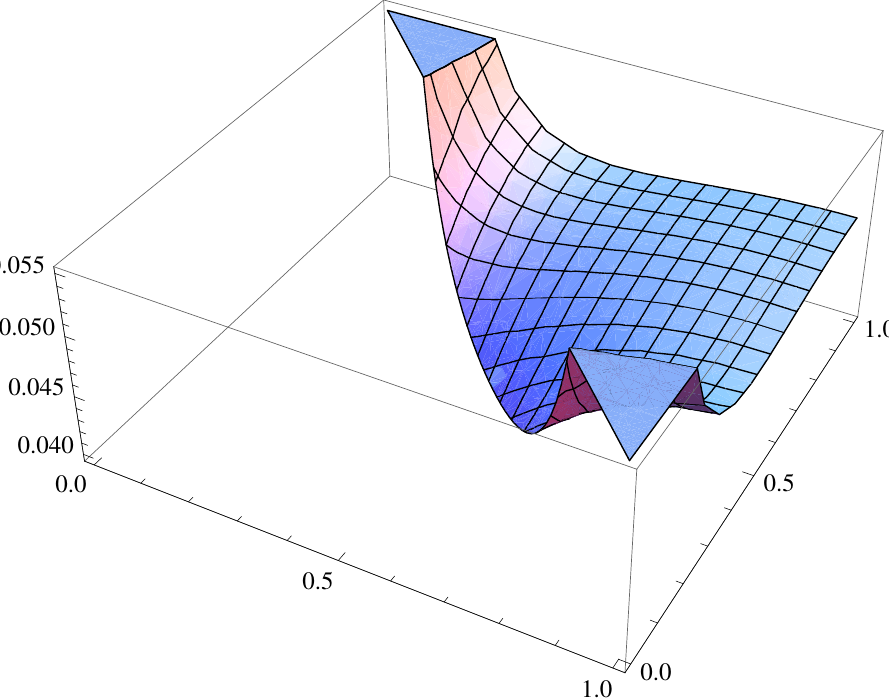} &
\includegraphics[width=0.3\linewidth]{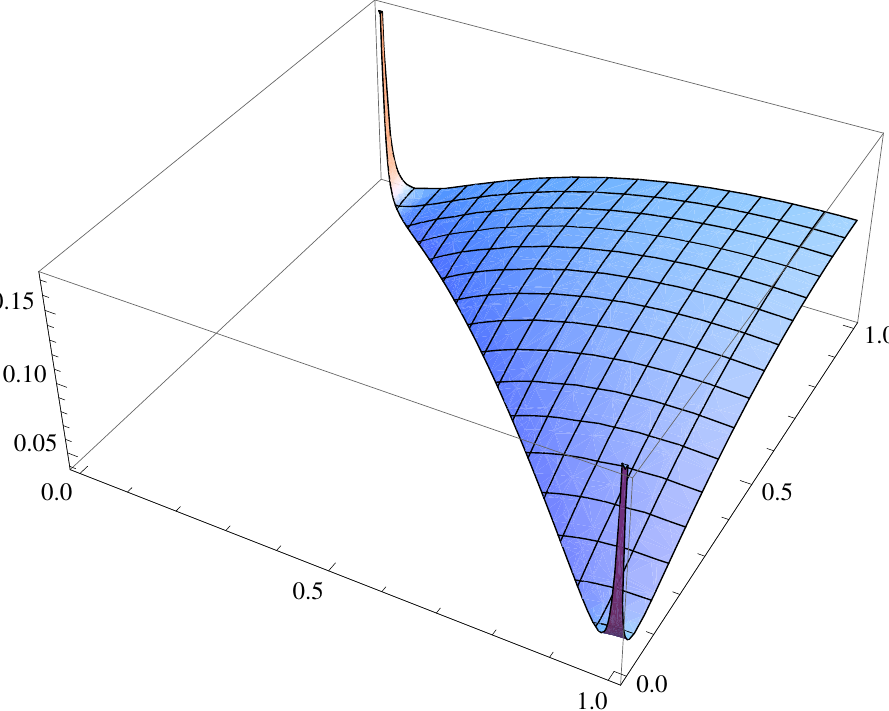} \\ 
\includegraphics[width=0.3\linewidth]{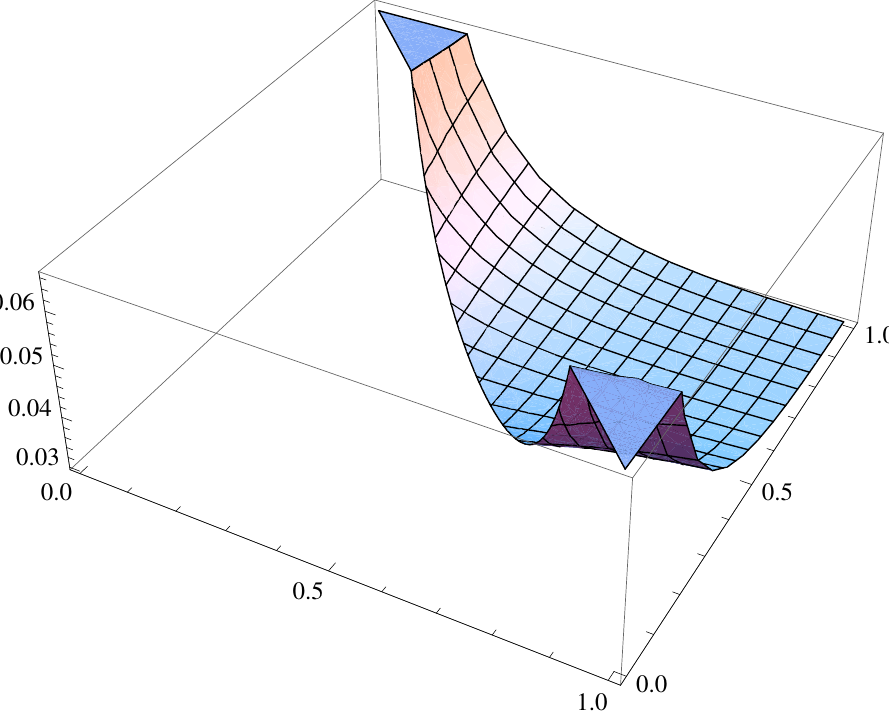} &
\includegraphics[width=0.3\linewidth]{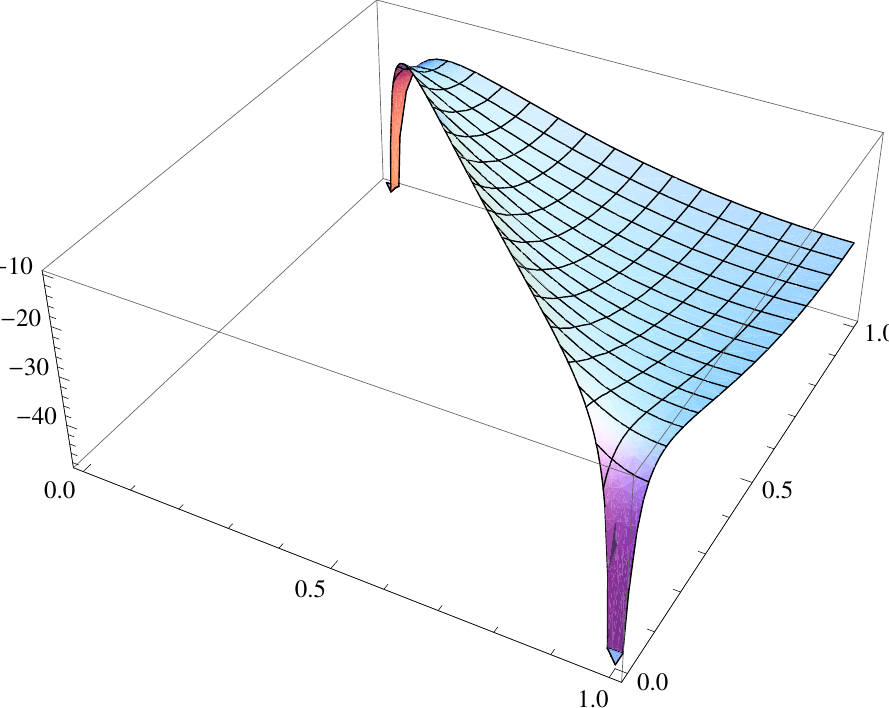} &
\includegraphics[width=0.3\linewidth]{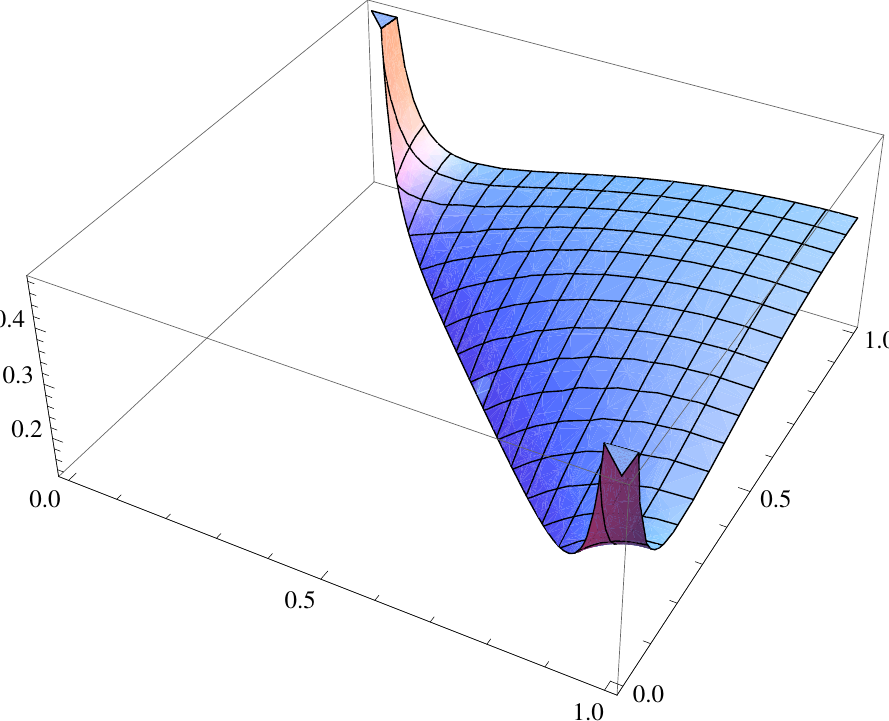} \\ 
\includegraphics[width=0.3\linewidth]{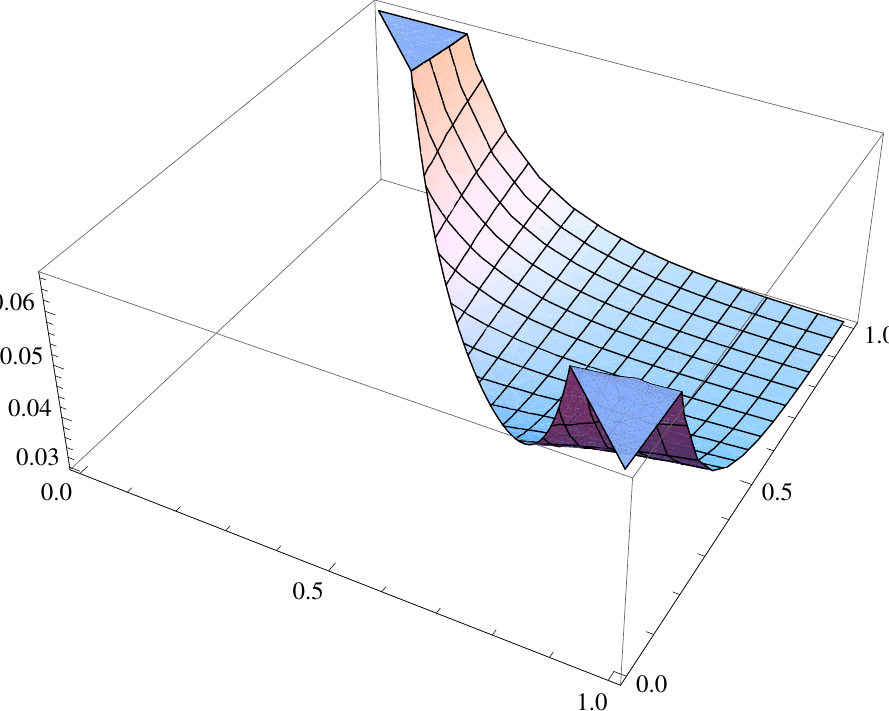} &
\includegraphics[width=0.3\linewidth]{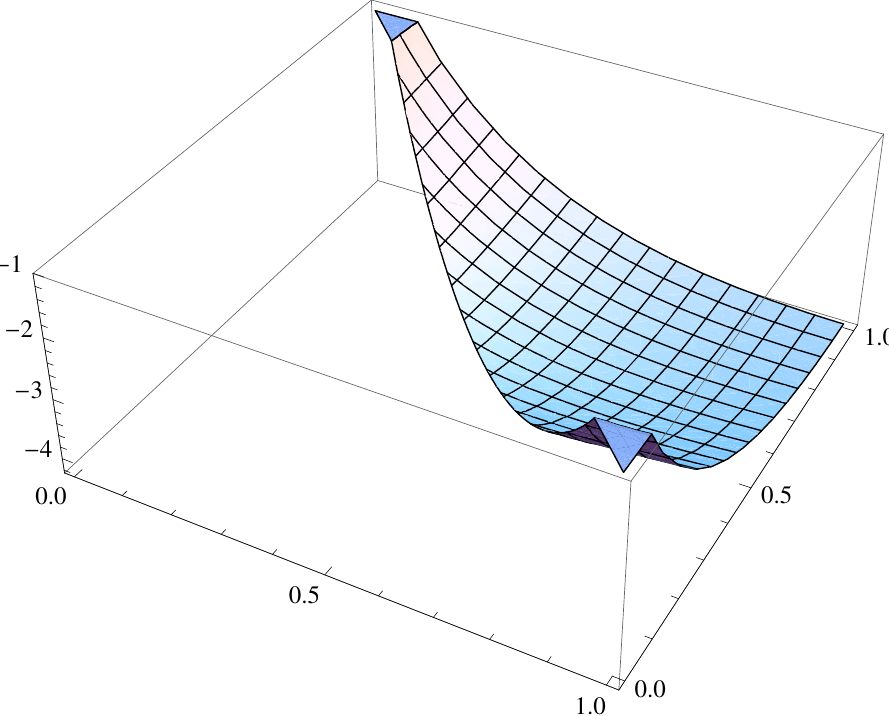} &
\includegraphics[width=0.3\linewidth]{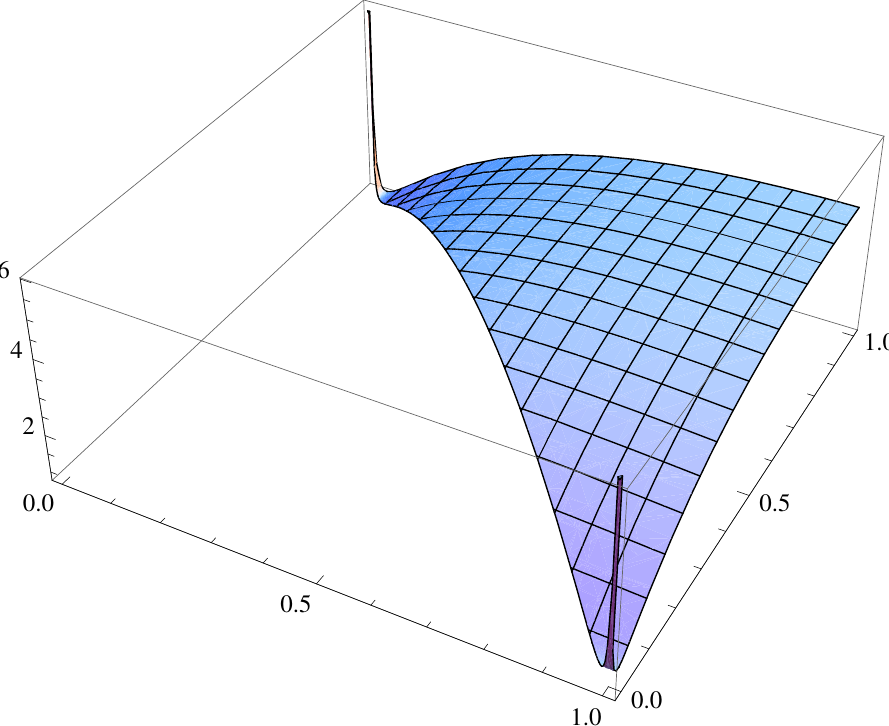}
\end{array}$
\end{center}
\caption{Here we plot the Non-Gaussian amplitude ${\cal A}(1, k_1 , k_2)/( k_1 k_2)$ disentangling effects from $\ep,\bar{c_s}$ and $f_X$. {\bf Top row: The effect of varying $\ep$.} The amplitude is plotted for $\ep = 0.01,\,0.1,\,0.3$ respectively from left to right. $\cb = 1$ and $f_X = 0.01$. {\bf Middle row: The effect of varying $\cb$.} The amplitude is plotted for $\bar{c}_s = 1,\,0.1,\,10$ respectively from left to right. $\ep = 0.01$ and $f_X = 0.01$.
{\bf Bottom row: The effect of varying $f_X$.} $f_X = 0.01,\,100,\,-100$ respectively from left to right. $\ep = 0.01$ and $\cb = 1$.}
\label{disentangledshapes}
\end{figure}

An interesting question to ask is therefore: {\it Considering deviations from slow-roll and exact scale-invariance, how robust is the statement that single field models with sizable non-Gaussianity have an equilateral shape?} In the slow-roll limit the cubic action~\eqref{action3} is dominated by its first three terms
\begin{eqnarray} \label{actionSR}
S_{slow-roll}&\sim&{M_{Pl}}^2\int {\rm d}t {\rm d}^3x \left\{
-a^3 \left[\Sigma\left(1-\frac{1}{c_s^2}\right)+2\lambda\right] \frac{\dot{\zeta}^3}{H^3} \right. \nonumber \\ &+&
\left. \frac{a^3\epsilon}{c_s^4}(\epsilon-3+3c_s^2)\zeta\dot{\zeta}^2 +  \frac{a\epsilon}{c_s^2}(\epsilon-2\es+1-c_s^2)\zeta(\partial\zeta)^2 + ... \right\}  ~,
\end{eqnarray}
This is visible from \eqref{amplitudes1} and \eqref{amplitudes2} as we can see there that the amplitudes corresponding to the remaining terms in the action are suppressed by powers of $\ep$. The leading order contributions to $\cA$ are therefore
\be
{\cal A}_{slow-roll} \sim {\cal A}_{\dot\zeta^3} + {\cal A}_{\zeta \dot\zeta^2} + {\cal A}_{\zeta(\partial \zeta)^2}.
\ee
A breaking of scale-invariance will then introduce a local component into the amplitude. This is shown in the left column of Figure~\ref{disentangledshapes}, where the amplitude is plotted for a slow-rolling near-canonical field (i.e. with $c_s \sim 1$ and $f_X \sim 0$) and spectral index $n_s = 0.96$. In fact it is known that, in the absence of large subhorizon interactions, the very squeezed limit is always dominated by a local shape contribution proportional to $n_s - 1$ \cite{malda,consistency}
\be \label{consistency}
\langle \zeta_{k_1} \zeta_{k_2} \zeta_{k_3} \rangle \sim -(n_s - 1) \delta^3(\sum_i {\bf k}_i)\frac{P_\zeta(k_1) P_\zeta(k_3)}{4 k_1^3 k_3^3}.
\ee
The middle and bottom row of Figure~\ref{disentangledshapes} then show how considering non-canonical fields with $c_s \ne 1$ and/or $|f_X| \gg 1$ can amplify $\cA$. The resulting shapes are predominantly equilateral. It is important to note that, whilst the local contribution from breaking of scale-invariance can be amplified by non-canonical kinetic terms, the only significant deviation from an equilateral shape is in the squeezed limit. The extreme squeezed limit is of course not observationally accessible, as it corresponds to considering modes with infinite wavelength. If the local contribution is suppressed outside the extreme squeezed limit, as in the case plotted in the middle row of Figure~\ref{disentangledshapes}, it is therefore possible that the observable section of the amplitude $\cA$ is consistent with it being purely equilateral, despite there being local contributions.

\begin{figure}[h] 
\begin{center}$
\begin{array}{cc}
\includegraphics[width=0.4\linewidth]{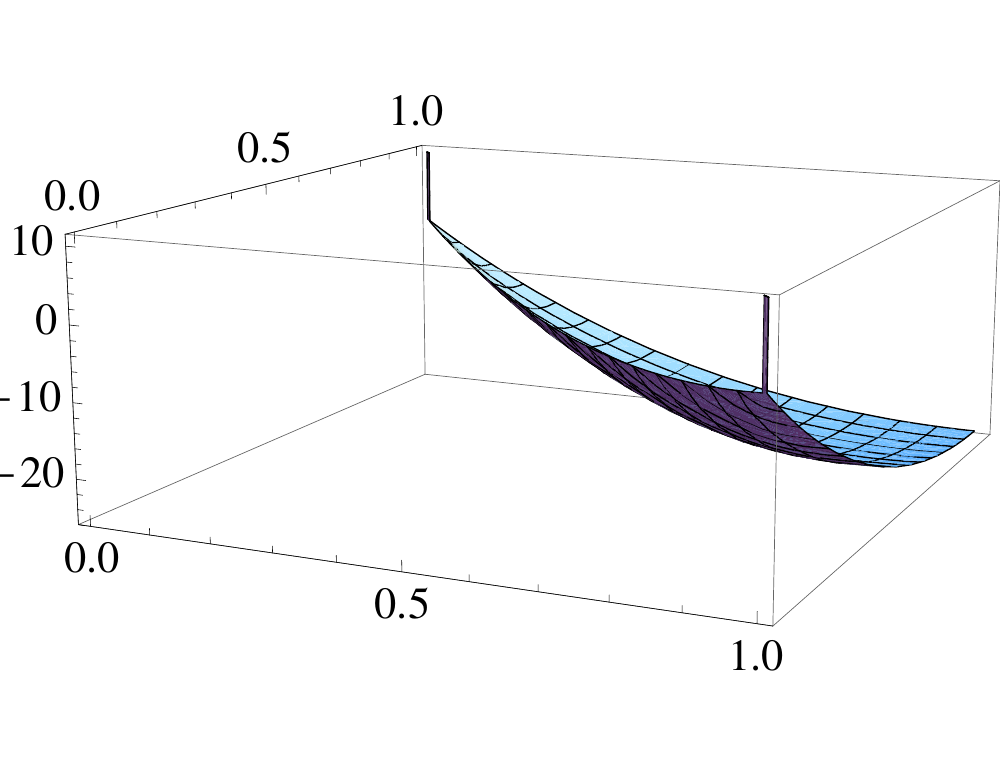} &
\includegraphics[width=0.4\linewidth]{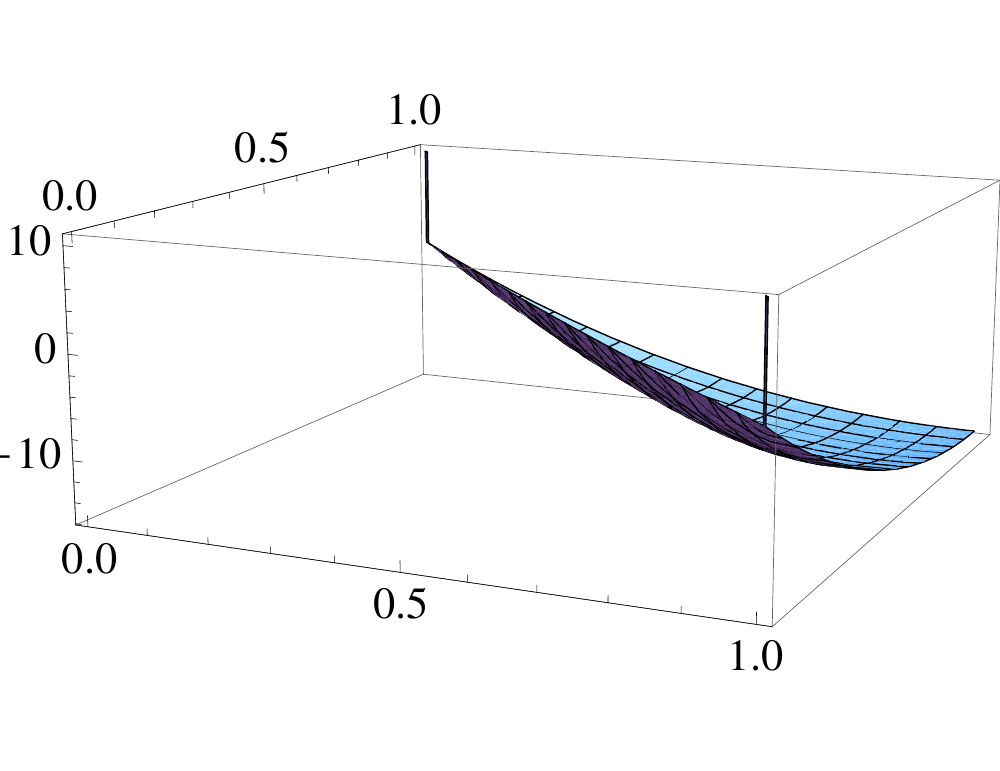} \\ 
\includegraphics[width=0.4\linewidth]{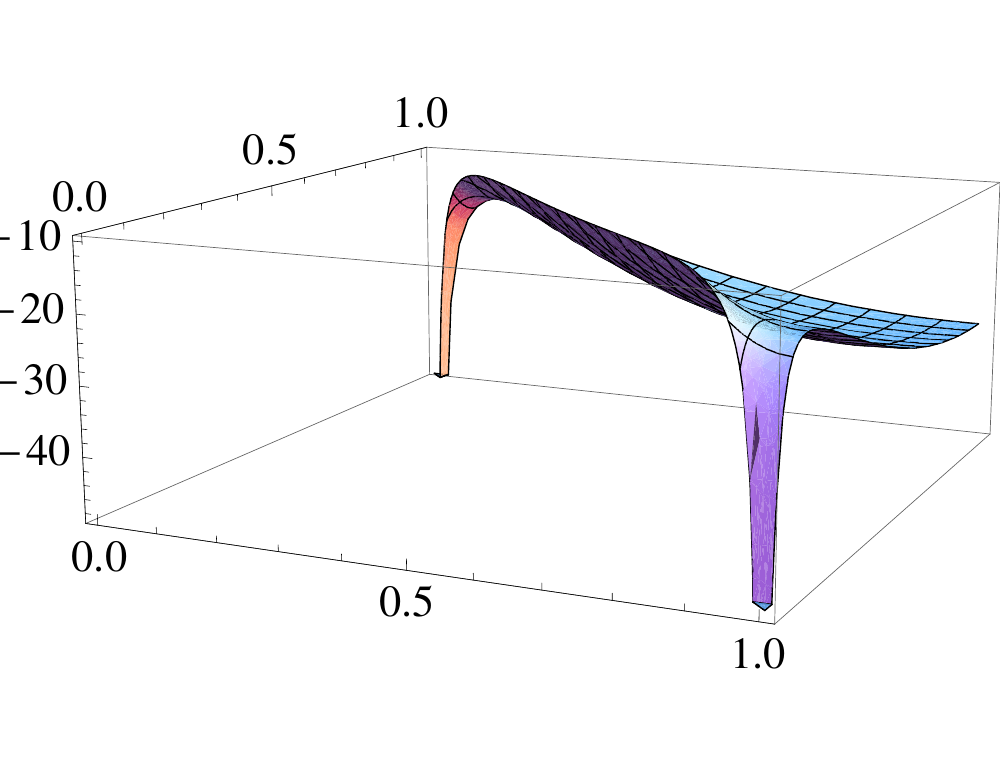} &
\includegraphics[width=0.4\linewidth]{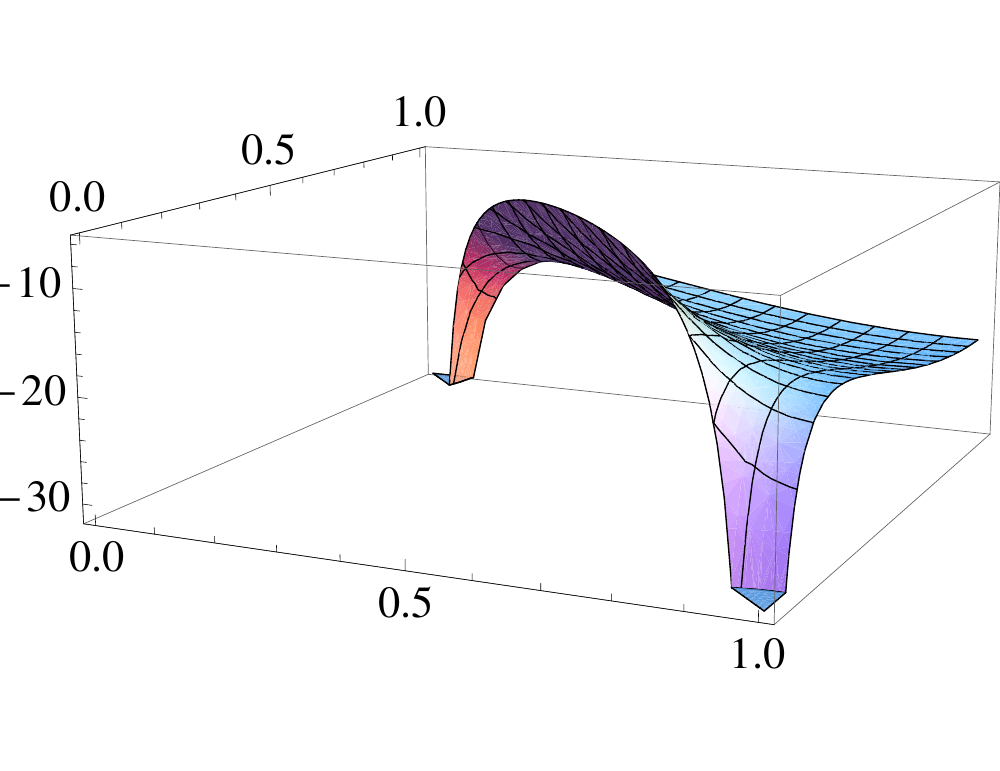} 
\end{array}$
\end{center}
\caption{
Here we plot the Non-Gaussian amplitude ${\cal A}(1, k_1 , k_2)/( k_1 k_2)$ for $f_X = 0.01$ and $c_s = 0.1$. Plots in the top and bottom rows are for $n_s = 1, 0.96$ respectively. Plots in the left and right columns are for $\ep = 0.001, 0.3$ respectively.}
\label{shapescs}
\end{figure}

\begin{figure}[h] 
\begin{center}$
\begin{array}{cc}
\includegraphics[width=0.4\linewidth]{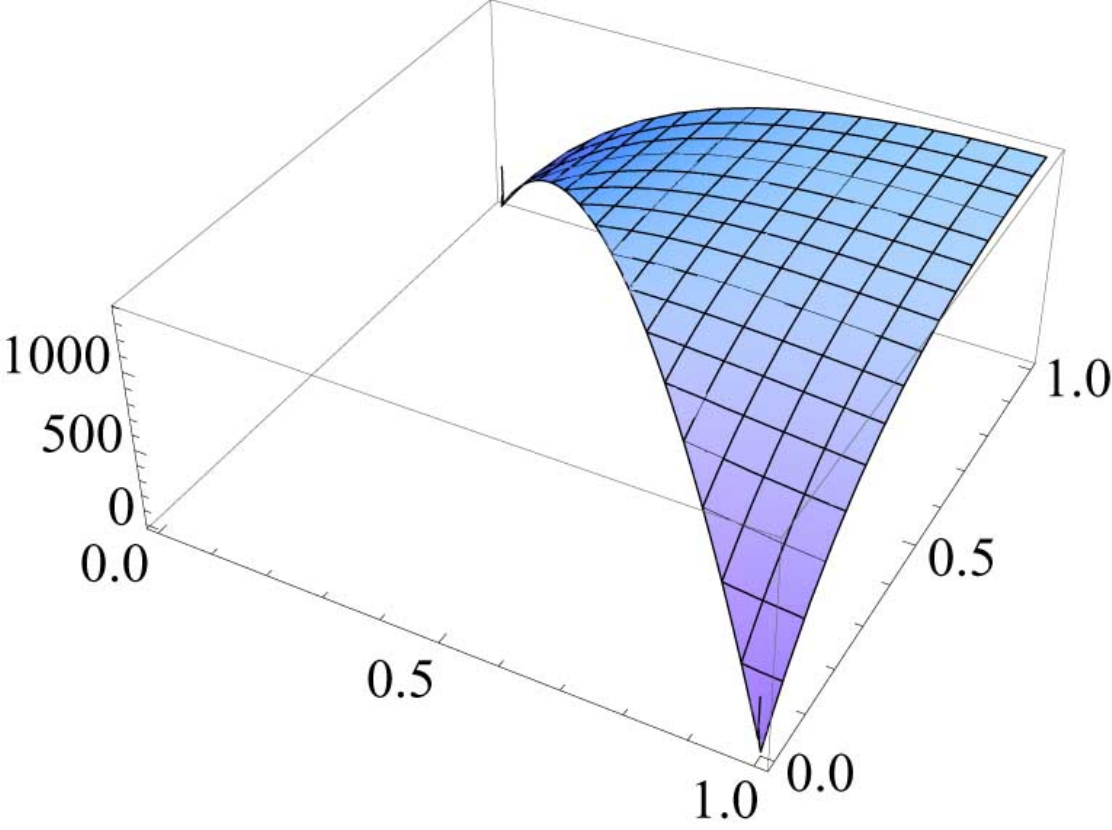} &
\includegraphics[width=0.4\linewidth]{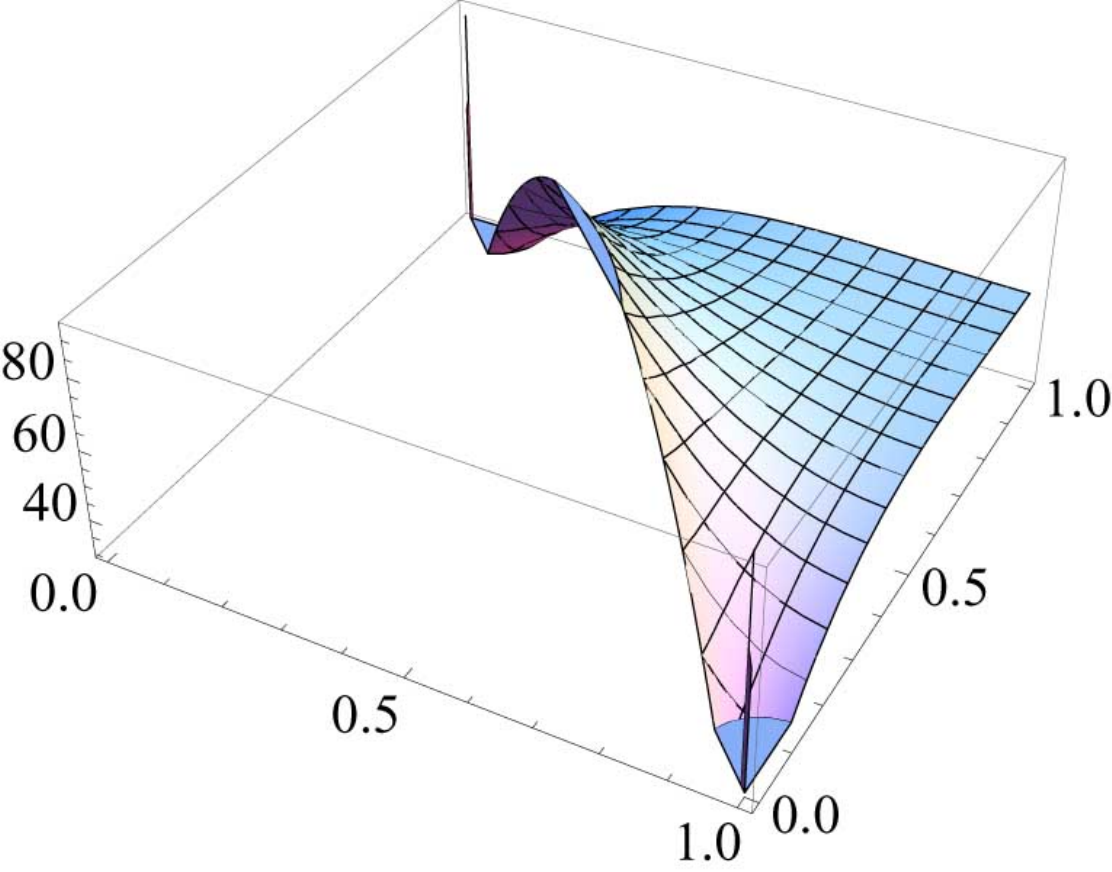} \\ 
\includegraphics[width=0.4\linewidth]{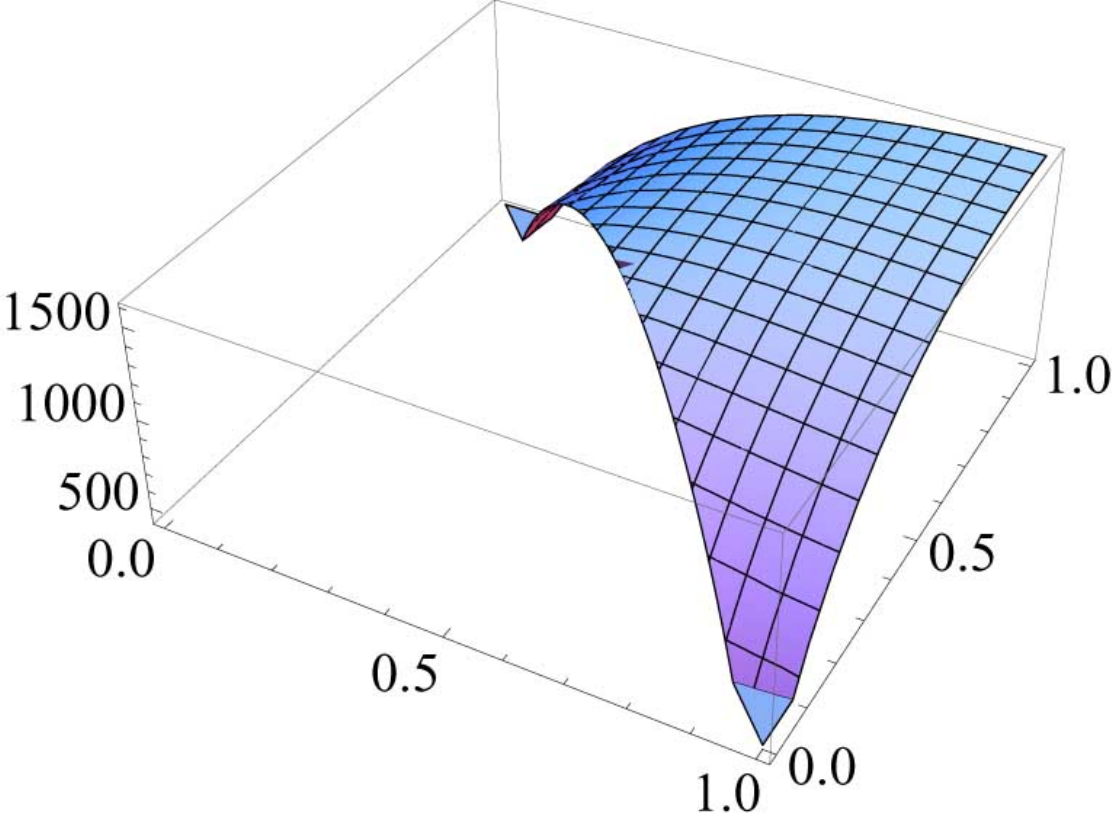} &
\includegraphics[width=0.4\linewidth]{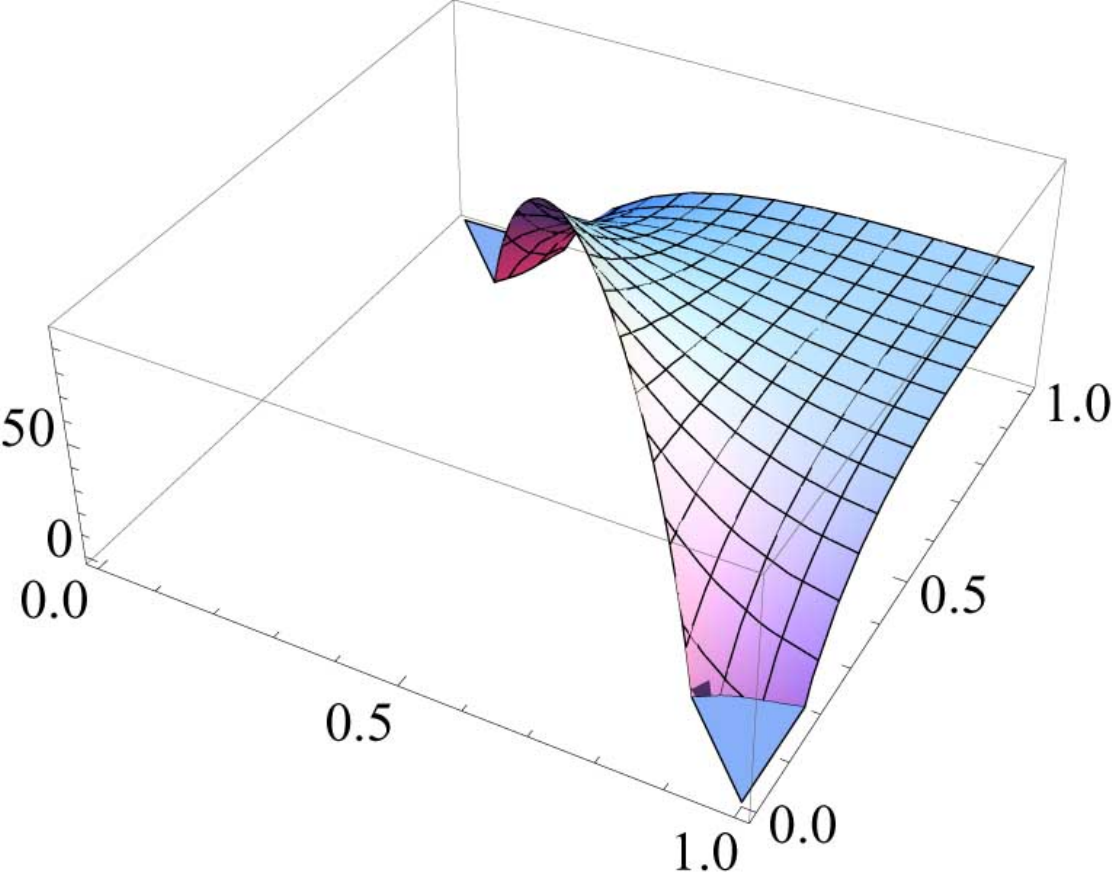} 
\end{array}$
\end{center}
\caption{Here we plot the Non-Gaussian amplitude ${\cal A}(1, k_1 , k_2)/( k_1 k_2)$ for $f_X = -100$ and $c_s = 0.05$. Plots in the top and bottom rows are for $n_s = 1, 0.96$ respectively. Plots for the left and right columns are for $\ep = 0.001, 0.3$ respectively.}
\label{shapescsfx}
\end{figure}

What happens once we violate the slow-roll conditions and consider cases with $\ep \sim {\cal O}(1)$ consistent with constraints derived above?  Firstly the remaining interaction terms in~\eqref{action3},$ {\cal A}_{\dot \zeta \partial \zeta \partial \chi}$ and $ {\cal A}_{\epsilon^2}$, which were previously suppressed by powers of $\ep$, now become relevant. Even more importantly the remaining terms receive large corrections from $\ep$. This is most easily shown by considering the case when non-Gaussianities are primarily sourced by $f_X \gg 1$. In this case we have
\be \label{srsupamp}
{\cal A}(f_X \gg 1) \sim {\cal A}_{\dot\zeta^3}\, \sim \frac{f_X -1}{2 {\bar c}_s^{2}} \left(\frac{k_1k_2k_3}{2 K^3}\right)^{n_s - 1} (\epsilon + \epsilon_s - 1) \cos\frac{\alpha_2 \pi}{2} \Gamma(3+\alpha_2)\frac{k_1^2k_2^2k_3^2}{K^3}.
\ee
As we already saw in the previous section~\eqref{fnlfx} this means the amplitude is suppressed by a factor of $\cos\frac{\alpha_2 \pi}{2} \Gamma(3+\alpha_2)$. Considering violations of slow-roll can therefore lead to interesting new phenomenology, as previously suppressed contributions become important. 

How can one understand the fast-roll suppression of the non-Gaussian amplitude as manifest in~\eqref{srsupamp} physically? In essence one can separate the effect of considering large $\epsilon,\epsilon_s$ on the amplitude into two categories. On the one hand the interaction Hamiltonian that follows directly from~\eqref{action3} is modified, since the ``coupling constants'' for individual interaction terms depend on $\ep,\epsilon_s$. The interaction terms one can ignore in the slow-roll limit~\eqref{actionSR} are all proportional to $\ep$, so that their effect linearly increases upon breaking slow-roll. This explains why in the canonical case with ignorable $f_x$ (as depicted in the left panel of figure~\ref{fnlfigure}) an increase in $\ep$ leads to an enhanced non-Gaussian signal. For in the canonical extreme slow-roll limit one is essentially considering a free field and interaction terms ``switch on'' as larger values of slow-roll parameters are considered. 

The second effect of large $\epsilon,\epsilon_s$, which lies at the root of the fast-roll suppression of $\cA$ discussed above, is a modification to the propagators~\eqref{propagator}. In other words the functional dependence of $\zeta$ on slow-roll parameters can then become important. The argument is analogous to the original one presented for the emergence of an equilateral shape for single field models. We recall that this argument invoked modes far inside the horizon oscillating so that their contributions to $\cA$ average out. Considering non-slow-roll propagators can now have a similar effect. For the appearance of suppression factors as shown in equations~\eqref{fnlfx} and~\eqref{srsupamp} is essentially due to a ``destructive interference'' of modes. This means cancellations between contributions to the amplitude $\cA$ can occur, since the oscillatory behaviour of $\zeta$ is modified.

Figures~\ref{shapescs} and~\ref{shapescsfx} show two concrete case studies to illustrate our results. Figure~\ref{shapescs} shows how the relative contribution of an orthogonal component is amplified by $\ep$ and how $n_s \ne 1$ induces a local component. The local component away from the extreme squeezed limit is also modified by $\ep \sim {\cal O}(1)$. The resulting amplitude gives rise to an intermediate shape with both local, equilateral and orthogonal parts. As the bottom right hand graph shows, this would be observationally distinguishable from the slow-roll and scale-invariant cases.

Figure~\ref{shapescsfx} illustrates the $\ep$-suppression of otherwise dominant contributions. The top left graph shows the amplitude for the given choice of parameters in the case of exact scale-invariance and slow-roll. A very large amplitude with $\fnl \sim 1000$ is produced, which violates present upper bounds on the level of non-Gaussianity. Deviating away from slow-roll has two important effects here. Firstly the fast-roll suppression moves the amplitude back within observational constraints. Therefore new regions in the parameter space $\{c_s,f_X\}$ open up. Secondly a delicate cancellation between terms can bring out contributions from otherwise suppressed orthogonal and local shapes even more here than in the example considered above. This is shown by the $\ep \sim {\cal O}(1)$ amplitudes in Figure~\ref{shapescsfx} peaking in the folded limit $2 k_1 \approx 2 k_2 \approx k_3$. A shape which cannot be decomposed into local and equilateral shapes alone, but requires a strong orthogonal component.

Violation of slow-roll and deviation from exact scale invariance therefore naturally lead to the generation of intermediate shapes with equilateral, local and orthogonal contributions. Hence single field models can produce a richer phenomenology than purely equilateral shape type non-Gaussianity. And whilst general statements about limits of the three-point function for single field models, such as~\cite{malda,consistency} remain true, such models have a more complex fingerprint when considering the full amplitude.

\subsection{The running $n_{NG}$} \label{running}

\begin{figure}[h]
\begin{center}$
\begin{array}{ccc}
\includegraphics[width=0.3\linewidth]{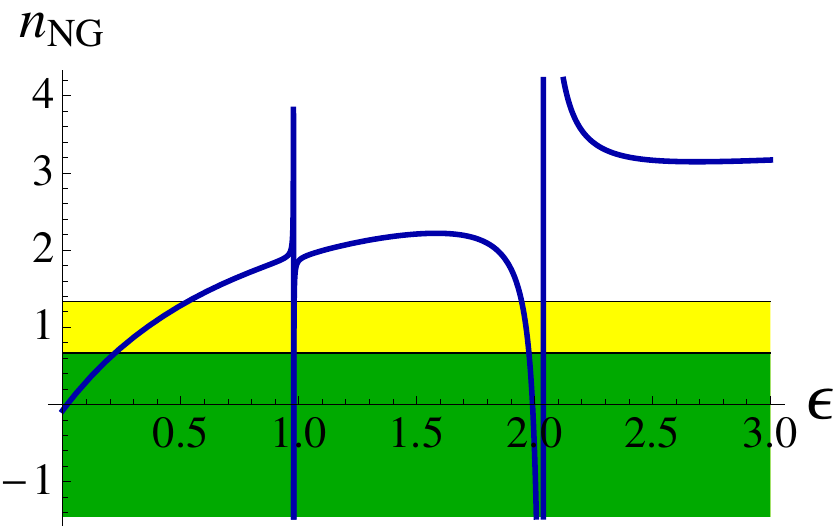} &
\includegraphics[width=0.3\linewidth]{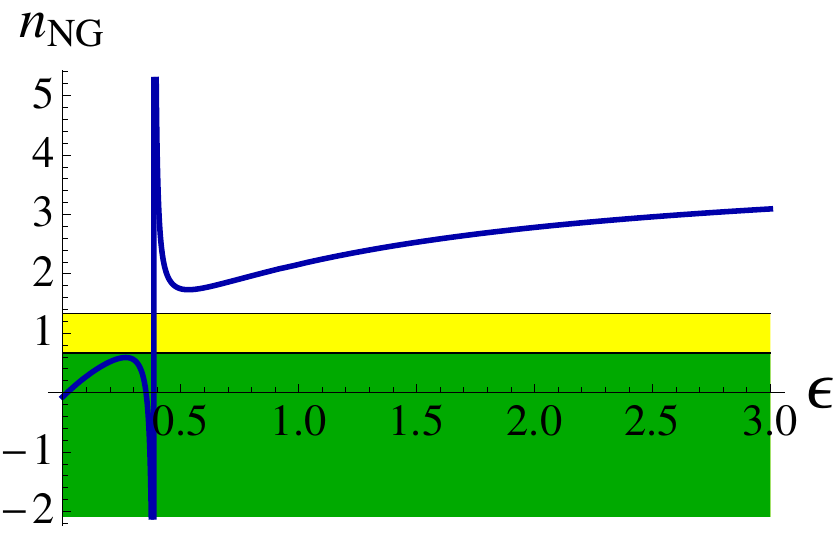} &
\includegraphics[width=0.3\linewidth]{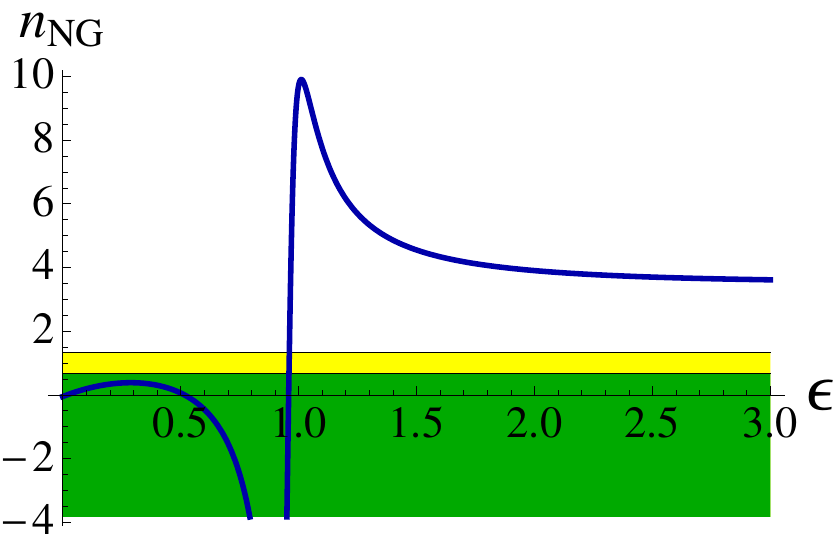} \\ 
\includegraphics[width=0.3\linewidth]{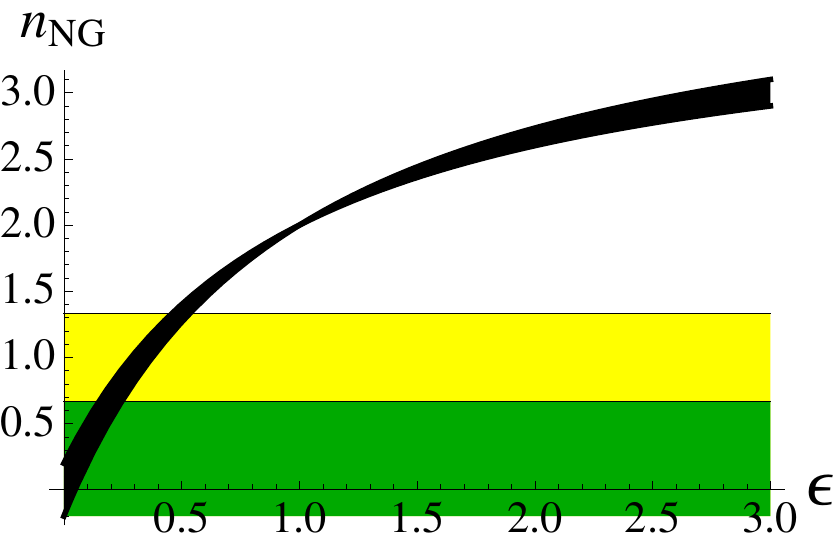} &
\includegraphics[width=0.3\linewidth]{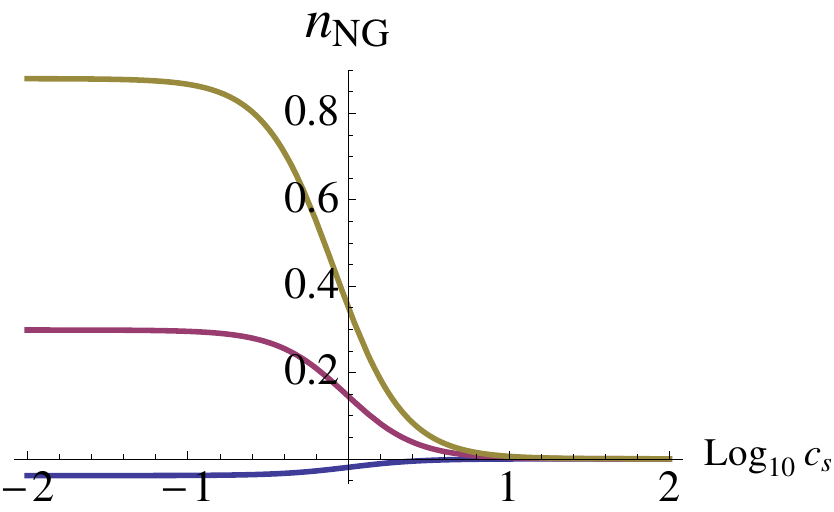} &
\includegraphics[width=0.3\linewidth]{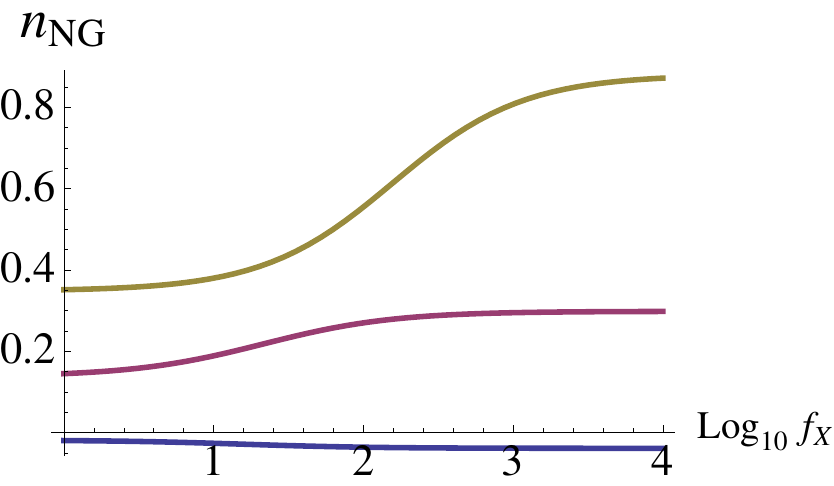}
\end{array}$
\end{center}
\caption{
{\bf Top row:} $n_{\rm NG}$ plotted against $\ep$ for $n_s = 0.96$. Green ($n_{\rm NG} < 2/3$) and yellow ($n_{\rm NG} < 4/3$) regions are those allowed by perturbative constraints on the level of Non-Gaussianity assuming $|f_{\rm NL}({\rm CMB})| \sim {\cal O}(100)$ and $|f_{\rm NL}({\rm CMB})| \sim {\cal O}(1)$ respectively. From left to right we plot: $\{ \cb = 0.1$,$f_X = 0.01 \}$ (left); $\{ \cb = 1$,$f_X = 100 \}$ (middle) ; $\{ \cb = 1$,$f_X = 10 \}$ (right).
{\bf Bottom left: } $n_{\rm NG}$ plotted against $\ep$ in the limit as $\cb \to 0$. The black-shaded region corresponds to values for $n_{\rm NG}$ for $n_s$ between $0.9$ and $1.1$. Almost identical plots are found for $\left| f_X \right| \gg 1$.
{\bf Bottom middle: } $n_{\rm NG}$ plotted against $\text{Log}_{10} (\bar{c}_s)$ for $n_s = 0.96$ and $\ep = 0.3,0.1,0.01$ from top to bottom respectively. $f_X = 1$.
{\bf Bottom right: } $n_{\rm NG}$ plotted against $\text{Log}_{10} (f_X)$ for $n_s = 0.96$ and $\ep = 0.3,0.1,0.01$ from top to bottom respectively. $\cb = 1$.
}
\label{nngplots}
\end{figure}

Having studied the shapes of non-Gaussianity at fixed $K$ in the previous section, we now investigate the running of $\cA$ with scale $K$~\cite{runningchen,runningconstraints,Byrnes:2010ft}. Introducing a dynamical speed of sound can in principle lead to a strongly scale-dependent non-Gaussian amplitude $\cA$. To see why we rewrite $\bar{c}_s$, the speed of sound at sound horizon crossing, in terms of the wavenumber $K$ (which, we remind ourselves, is defined as $K = k_1 + k_2 + k_3$)
\be
\bar{c}_s \sim K^{\frac{- \epsilon_s}{\epsilon_s + \epsilon - 1}}.
\ee
Different scales $K$ will therefore ''see`` a different $\bar{c}_s$ upon crossing the horizon. To quantify this difference we follow~\cite{runningchen} in defining a spectral index for $f_{\rm NL}$ as
\be
n_{\rm NG} - 1 \equiv \frac{{\rm d}\ln |f_{\rm NL}|}{{\rm d}\ln K}\,,
\ee
where we evaluate the running $n_{NG}$ at a fixed point of the amplitude as measured by $\fnl$ to separate effects from the running and shape. Comparison with~\eqref{amplitudes1} and~\eqref{amplitudes2} shows that we can always schematically express $f_{\rm NL}$ as $f_{\rm NL} = {\cal C}_1(n_s,\ep,f_X) (1 - {\cal C}_2(n_s,\ep,f_X) \bar{c}_s^{-2}$), where ${\cal C}_1$ and ${\cal C}_2$ are functions of the slow-roll parameters  ${\cal C}_i (\ep,n_s,f_X)$ or equivalently ${\cal C}_i (\ep,\eps,f_X)$ and are given exactly in appendix~\ref{generalfnl}. As such we obtain
\be
\ln |f_{\rm NL}| = \ln |{\cal C}_1| + \ln  |1 - {\cal C}_2 \bar{c}_s^{-2}|.
\ee 
The non-Gaussian tilt $n_{NG}$ is therefore given by
\bea
n_{NG} - 1 &=& \frac{2 + 2 \ep (-3 + n_s) - 2 n_s}{1 + \ep}  \frac{{\cal C}_2 (\ep,\eps,f_X)}{\bar{c}_s^2 - {\cal C}_2 (\ep,\eps,f_X)} \nonumber \\
&=& \frac{-2 \eps}{\eps + \ep - 1} \frac{{\cal C}_2}{\cbt - {\cal C}_2}.\label{nnggeneral}
\eea

In the small speed of sound limit and for an exactly scale-invariant 2-point function in the $\ep > 0$ regime the 3-point function acquires a blue tilt~\cite{kp}
\be
n_{\rm NG} - 1  \sim   \frac{4\epsilon}{\epsilon + 1}.
\ee
This is because exact scale-invariance is associated with $\eps = -2 \ep$ and by requiring $\ep > 0$ we have prohibited solutions with a phantom equation of state $w < -1$. However, for slightly tilted 2-point functions the behavior of $n_{NG}$ is not as simple as we will see. In fact for $n_s = 0.96$ and in the slow-roll limit one obtains a slightly red tilted $n_{NG}$.

Can we put constraints on the value of the tilt of the 3-point function?
In principle levels of non-Gaussianity are observable for all scales running from CMB scales ($k^{-1} \sim 10^3 {\rm Mpc}$) down to galactic scales ($k^{-1} \sim 1 {\rm Mpc}$)~\cite{kp,runningconstraints,lssng}. In terms of the scale $K$ this therefore corresponds to $K_{\rm gal}/K_{\rm CMB} \simeq 10^3$. For the non-Gaussian tilt this means
\be
f_{\rm NL} ({\rm CMB})\ \approx \ 10^{-3(n_{\rm NG} - 1)} f_{\rm NL} ({\rm Gal})\,.
\ee
Blue/red tilted 3-point functions thus correspond to larger/smaller non-Gaussian amplitudes on smaller scales. 
In order for a perturbative treatment to be applicable the non-Gaussian contribution to $\zeta$ clearly must be much smaller than the Gaussian part $f_{\rm NL} \ll \zeta^{-1} \simeq 10^5$ for all observable scales. Present observational constraints on CMB scales give $f_{\rm NL} (\text{CMB}) \lesssim {\cal O} (100)$. In case of a scale-invariant or red-tilted 3-point function, where the amplitude is largest for large scales, a perturbative treatment is valid all the way down to the smallest scales. However, for blue-tilted 3-point functions one can find interesting constraints on $n_{NG}$ and hence on the slow-roll parameters. Specifically we have
\bea
n_{\rm NG} - 1 \lesssim 2/3 \hspace{.5cm} \text{for} \hspace{.5cm}  |f_{\rm NL}({\rm CMB})| \sim {\cal O}(100) \nonumber \\
n_{\rm NG} - 1 \lesssim 4/3 \hspace{.5cm} \text{for} \hspace{.5cm}  |f_{\rm NL}({\rm CMB})| \sim {\cal O}(1). 
\eea

In Figure~\ref{nngplots} we show how these bounds translate into constraints on the parameter space $\left\{ n_s,\ep,\cb\right\}$.
The top row shows that the precise bounds on $\ep$ resulting from $n_{NG}$ are in fact model-dependent. The left two graphs show typical plots for large $f_X$ and small $c_s$. Very roughly one can see that both scenarios require $\ep \lesssim 0.5$ to satisfy the weaker bound for $|f_{\rm NL}({\rm CMB})| \sim {\cal O}(1)$. Whilst this constraint is sufficient for the stronger bound  $|f_{\rm NL}({\rm CMB})| \sim {\cal O}(100)$ in the case of large $f_X$, this is not the case for a large $\fnl$ sourced by $c_s < 1$. There we find $\ep \lesssim 0.25$ as a requirement. The right graph finally shows that models can be tuned to satisfy bounds on $n_{NG}$ for all values in the range $0 < \ep < 1$. Interestingly there are also regions  with $\ep > 1$ where $n_{\rm NG}$ is well-behaved and the constraints are satisfied. This may be of importance in non-inflationary model-building. To summarize, we therefore find that the bounds on $\ep$ derived from $n_{NG}$ are model-dependent and significantly weakened compared to those derived in~\cite{kp}. 

However, in the limits as $f_X \to \infty$ or very small $c_s \to 0$ these graphs show the same behavior asymptotically. This is shown in the bottom left graph, which furthermore illustrates the dependence of bounds on $n_s$. One can read off $\ep \lesssim 0.5$ and $\ep \lesssim 0.25$ as approximate bounds for $|f_{\rm NL}({\rm CMB})| \sim {\cal O}(1)$ and $|f_{\rm NL}({\rm CMB})| \sim {\cal O}(100)$ respectively. If we focus on the stronger bound for $|f_{\rm NL}({\rm CMB})| \sim {\cal O}(100)$, it is also interesting to note that, whilst the constraint for red-tilt with $n_s \sim 0.96$ is $\ep \lesssim 0.25$, for a blue tilt we obtain the stronger bound $\ep \lesssim 0.15$ here. 

For simplicity the explicit examples and plots we provide throughout the paper are therefore chosen to satisfy bounds
\be
\ep \lesssim 0.3  \quad\quad,\quad\quad \fnl(\text{CMB}) \lesssim 100.
\ee
This ensures constraints from $n_{NG}$ are met for models with slightly red $n_s$. However, we stress that choosing larger values of $\fnl$ (especially of the equilateral type) at the expense of reducing $\ep$ or fine-tuning models to loosen bounds on $\ep$ is possible in accordance with the expressions derived above.

The bottom middle graph then reiterates the way in which a large $c_s$ suppresses $n_{\rm NG}$. Strong constraints on the slow-roll parameter $\ep$ are therefore only obtained for small speeds of sound. The graph also nicely shows how we can interpolate between red and blue tilts of the three-point correlation function by choosing suitable background variables. Specifically the line plotted for $\ep = 0.01$ displays a red tilt. The bottom right graph shows the analogous plot of $n_{NG}$ vs $f_X$. 

Constraints from $n_{NG}$ can therefore be used to put bounds on $\ep$. These bounds can become important here, since considering non-slow-roll models naturally gives rise to  a large running $n_{NG}$ of non-Gaussianity with scale, which are consequently a smoking-gun signature of such fast-roll models. Together with measurements of the spectral tilt $n_s$ of the 2-point correlation function, measuring $n_{NG}$ would allow a direct measurement of $\ep$ for single field models as considered here. Again we reiterate that we use ``fast-roll'' as a description for an evolution which violates the ``slow-variation'' condition $\ep \ll 1$ and not just as a reference to properties of the potential $V(\phi)$ (see discussion in section~\ref{singlefield}).

\section{Concrete model examples}\label{concrete}

\begin{figure}[h] 
\begin{center}$
\begin{array}{ccc}
\includegraphics[width=0.3\linewidth]{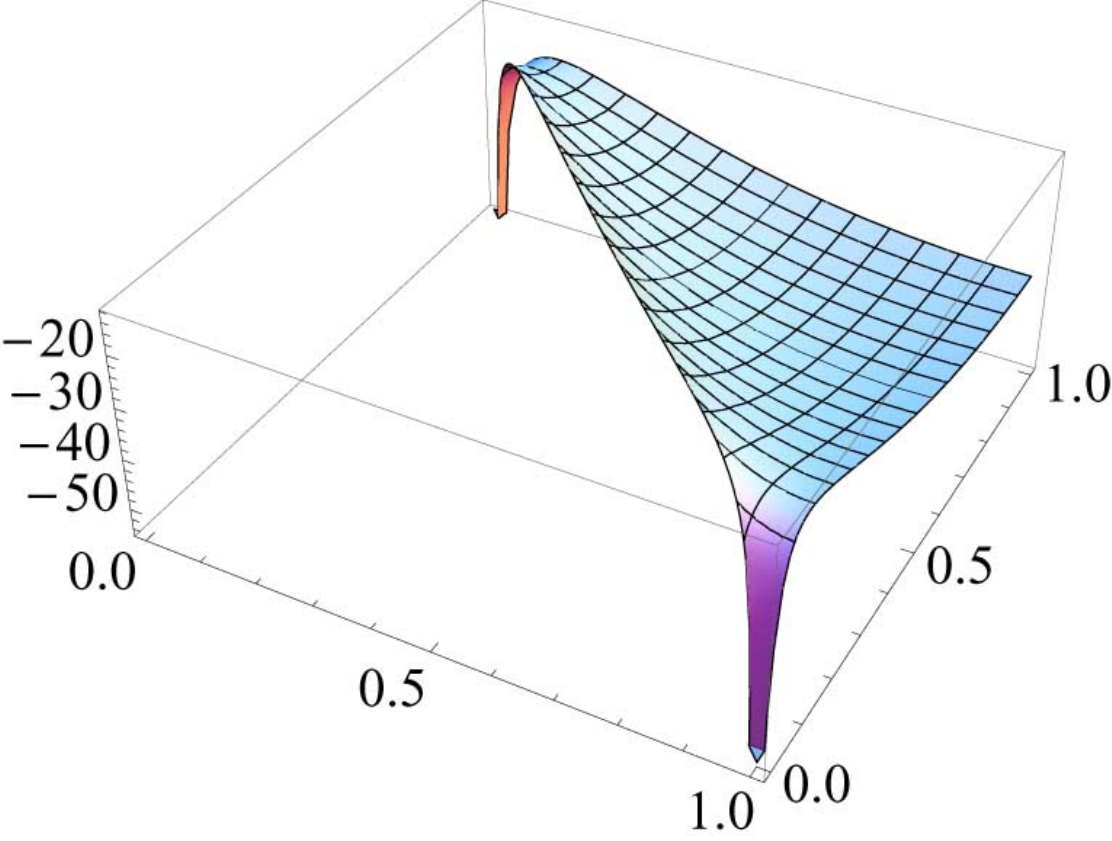} &
\includegraphics[width=0.3\linewidth]{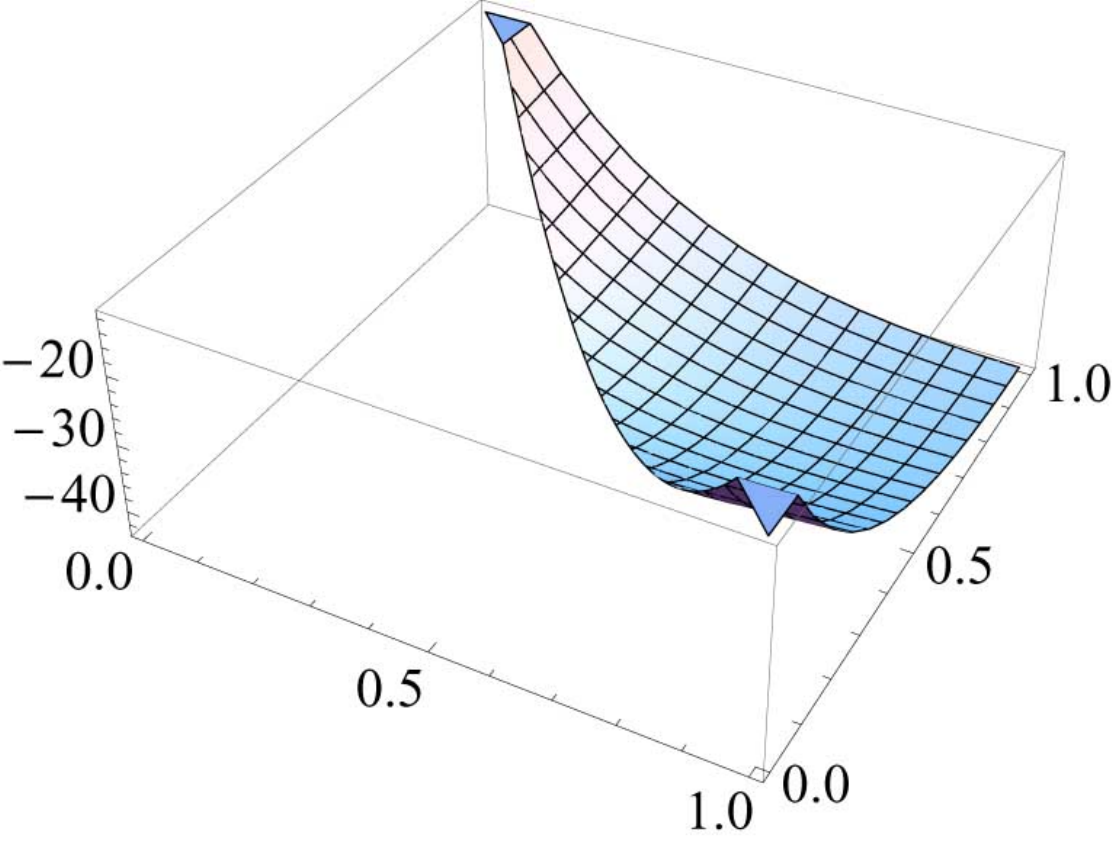} &
\includegraphics[width=0.3\linewidth]{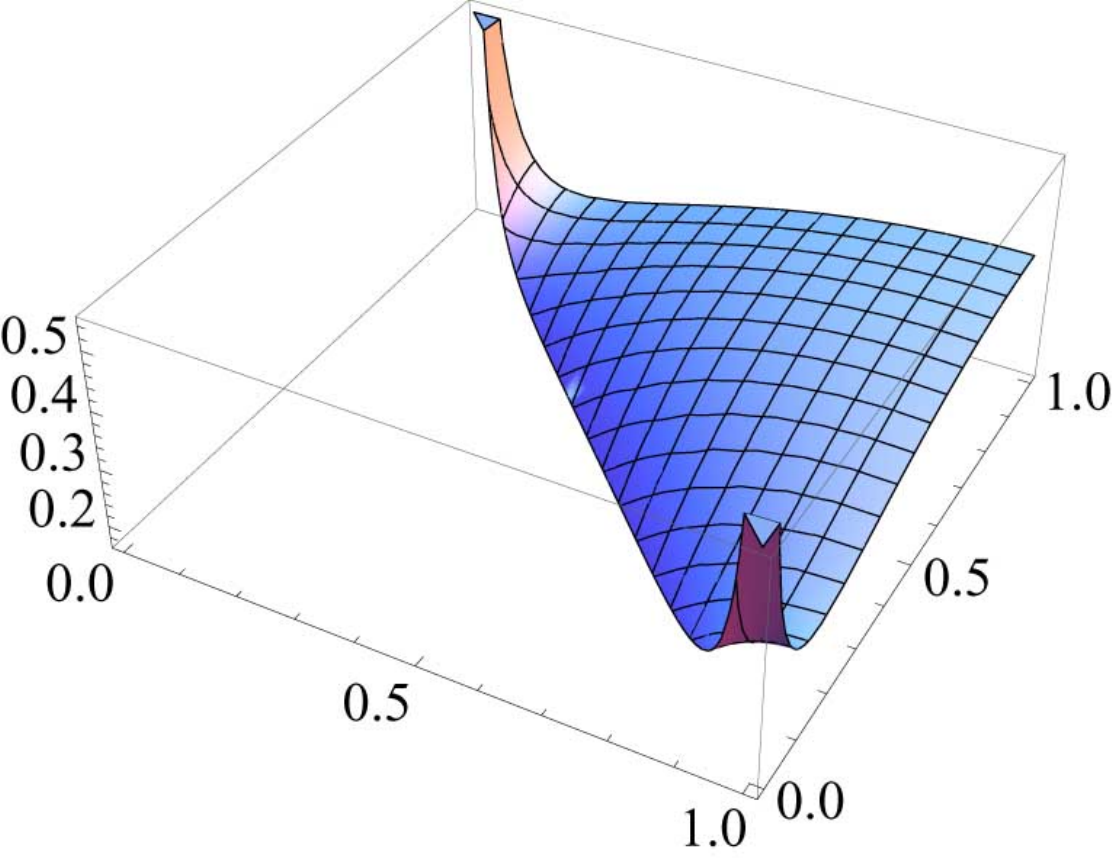} \\ 
\includegraphics[width=0.3\linewidth]{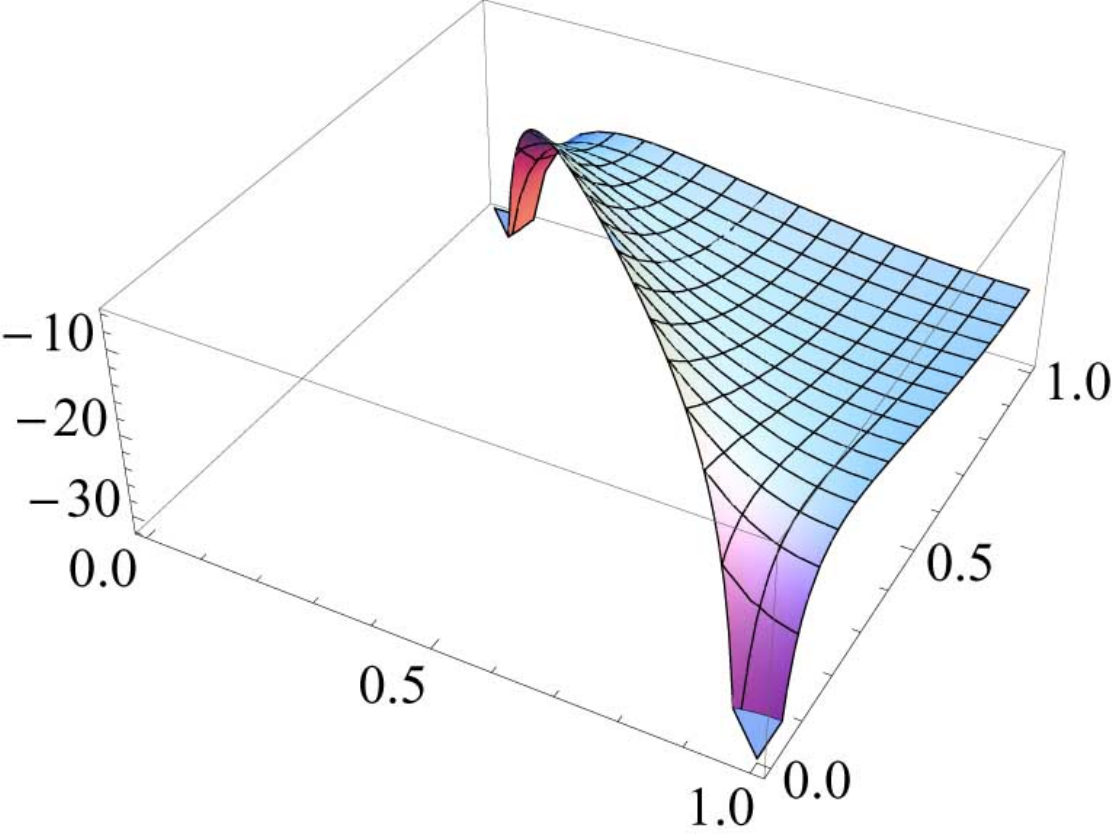} &
\includegraphics[width=0.3\linewidth]{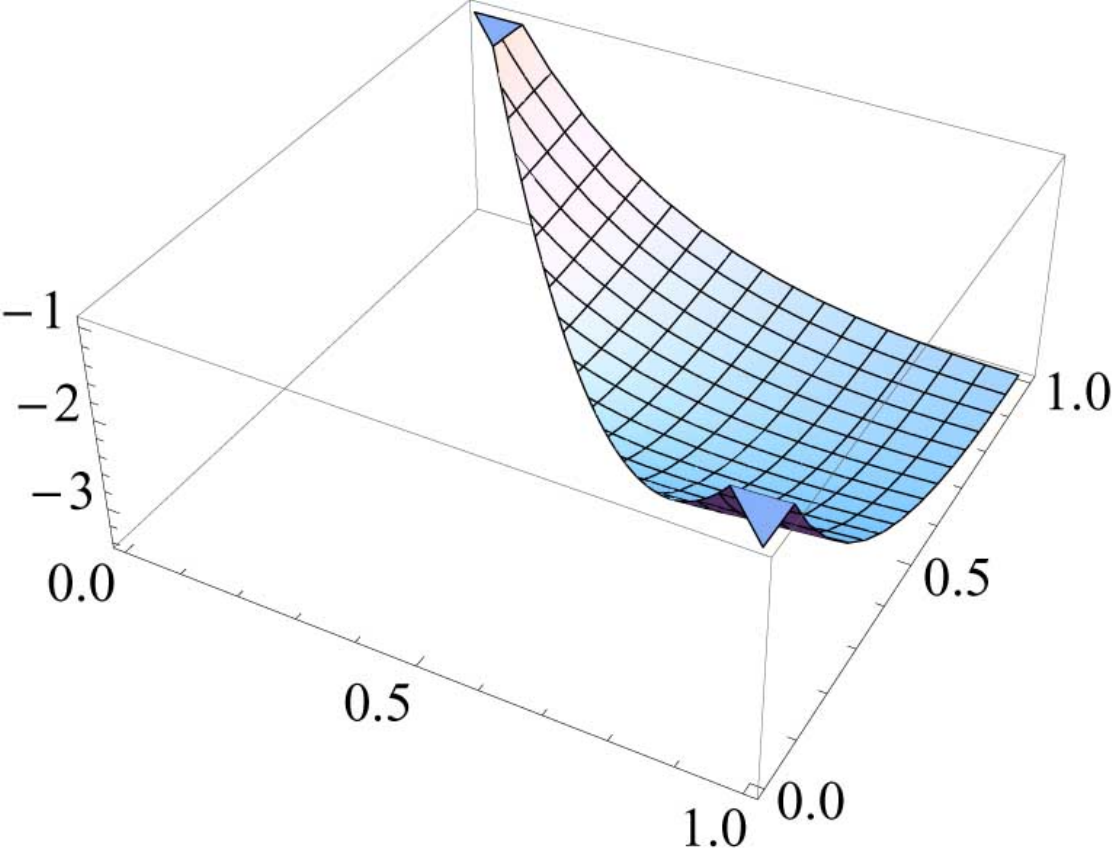} &
\includegraphics[width=0.3\linewidth]{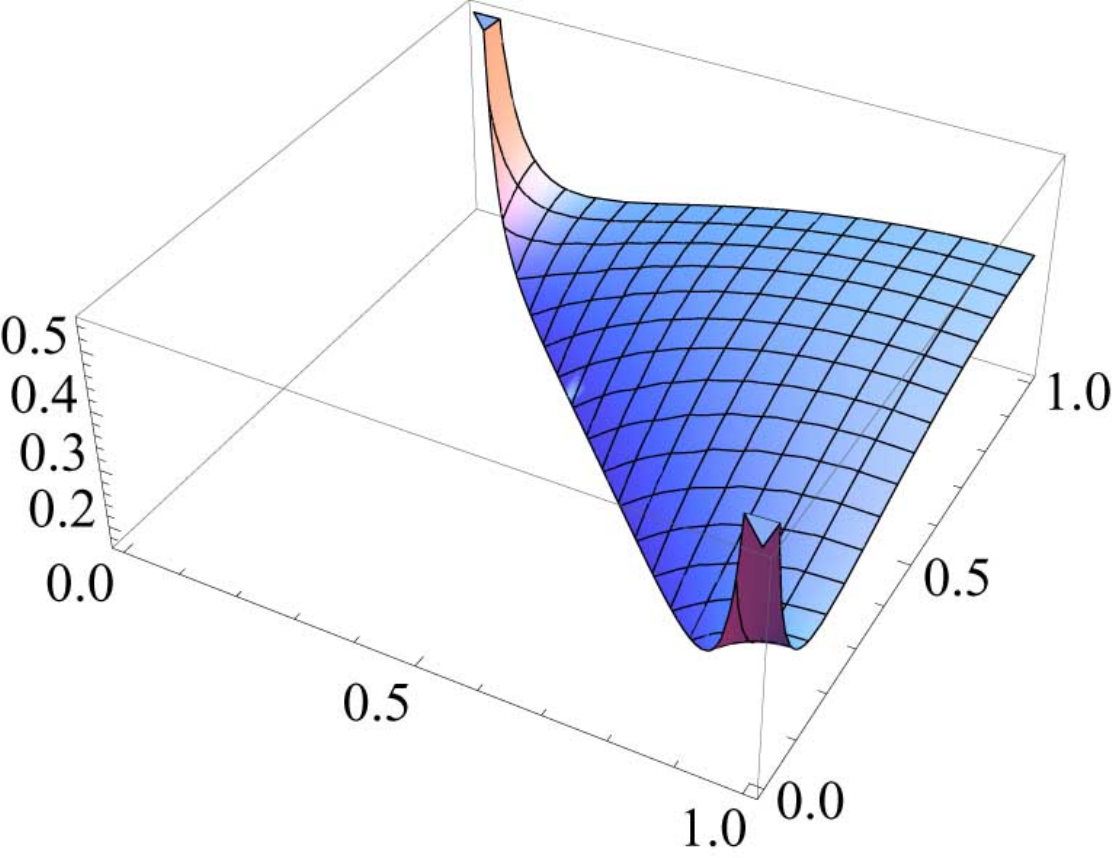}
\end{array}$
\end{center}
\caption{Here we plot the Non-Gaussian amplitude ${\cal A}(1, k_1 , k_2)/( k_1 k_2)$ for three specific models. $n_s = 0.96$ throughout. {\bf Left column: DBI inflation.} The amplitude is plotted for $\ep = 0.01$(top)and $\ep = 0.3$ (bottom) respectively. {\bf Middle column: Inflation with $\left|\frac{\lambda}{\Sigma}\right| \gg 1$.} The amplitude is plotted for $\ep = 0.01$(top)and $\ep = 0.3$ (bottom) respectively. $f_X = 1000$ and $\cb = 1$. {\bf Right column: Bimetric theories with large $c_s$.} The amplitude is plotted in the limit as $\cb \to \infty$ and for $\ep = 0.01$(top)and $\ep = 0.3$ (bottom) respectively. This confirms that $\cA$ is not sensitive to $\ep$ for bimetric theories in the large speed of sound limit.}
\label{concreteshapes}
\end{figure}

In the previous sections we derived general expressions for the levels of non-Gaussianity exhibited by single field models~\ref{general} without assuming slow-roll and discussed their associated phenomenology. Here we consider three concrete model implementations, writing down explicit actions for each model and showing how departure from slow-roll and scale-invariance affects their non-Gaussian signatures.

For single field models with large non-Gaussianity there are two interesting limits. From \eqref{oom} we have
\be
\fnl \sim {\cal O}(c_s^{-2}) + {\cal O}(\frac{\lambda}{\Sigma}),
\ee
so that models with $c_s \ll 1$ and $\left| \lambda/\Sigma \right| \gg 1$ will lead to large non-Gaussianities. We also considered another case with a smaller, but potentially detectable, non-Gaussian signal $\fnl \sim {\cal O} (1)$, when $c_s \gg 1$. The three concrete examples we provide are therefore examples of models with these signatures: ({\bf I}) DBI inflation ($c_s \ll 1$), ({\bf II}) Models with $\lambda / \Sigma \gg 1$ and ({\bf III}) Bimetric theories of structure formation ($c_s \gg 1$).

\subsection{Fast-roll DBI inflation} \label{dbisection}

In DBI inflation~\cite{Alishahiha:2004eh} a 3 + 1 dimensional brane moves in warped extra dimensions, giving rise to an effective 4D theory with action~\eqref{general}, where
\be P(X,\phi) =
-f^{-1}(\phi)\sqrt{1-2f(\phi)X}+f^{-1}(\phi)-V(\phi)\,. \label{DBI}
\ee
$f(\phi)$ is the so-called warp factor and is constrained to be positive by the signature of the space (in this paper we use the convention $-+++$). We only consider the 4-D effective field theory defined by~\eqref{DBI}. If the full dynamics of a particular higher dimensional implementation are considered, far stronger constraints (e.g. from gravitational waves~\cite{ian}) can be obtained. The speed of sound in DBI models is given by
\be
c_s = \sqrt{1 - 2 f(\phi)X}\label{dbics}
\ee
which is the inverse of the Lorentz boost factor $\gamma$, so $c_s \ll 1$ since the inflaton rolls ultra-relativistically.
Importantly one also finds
\be
\lambda = \frac{H^2 \ep}{2 c_s^4}(1-\ct) \quad\quad,\quad\quad f_X^{\rm DBI} = 1 - \ct,
\ee
so that the otherwise typically dominant ${\cal A}_{\dot\zeta^3}$ contribution vanishes, and the amplitude consequently becomes independent of parameter $f_X$, resulting in:
\be
\fnl^{\rm DBI} \sim {\cal O}(c_s^{-2}).
\ee
DBI inflation is the leading example of a single field model with large non-Gaussianity to be found in the literature, and it is associated with an equilateral shape $\cA$~\cite{chks,chenreview,largeng}. Figure~\ref{concreteshapes} confirms this for the case of slow roll $\ep \ll 1$. As expected the equilateral shape receives a small local correction from $n_s \ne 1$, which becomes relevant in the squeezed limit. However, just as for the more general single field case considered in section~\ref{shapes}, we find that departure from slow-roll suppresses the overall amplitude and widens the peaks in the squeezed limit. This happens as a consequence of introducing orthogonal and local contributions, thus potentially making them observationally accessible.

\subsection{Fast-roll inflation with $\left|\frac{\lambda}{\Sigma}\right| \gg 1$}

We now build an explicit action for a model realizing $|{\lambda}/{\Sigma}| \gg 1$. First we consider a number of constraints that  can be placed on the differential properties of $P(X,\phi)$. We can write slow-roll parameter $\ep$ as
\be
\ep = \frac{3}{2 - \frac{P}{X P_{,X}}}\; .
\ee
In order to satisfy the null energy condition $p + \rho \ge 0$ and $1 \le \ct < 0$ one must require, respectively:
\be
2 X P_{,X} > 0. \quad\quad,\quad\quad P_{,XX} > 0,
\ee
as can be seen from the expression for the speed of sound
\be
c_s^2 = \frac{P_{,X}}{\rho_{,X}}= \frac{P_{,X}}{P_{,X}+2X P_{,XX}}.
\ee
Observational bounds on parameters such as $n_s,r,n_t$ therefore only constrain the first two derivatives $X P_{,X}$ and $X^2 P_{,XX}$~\cite{beanrecon}.
Expressing $\lambda$ and $\Sigma$ in terms of derivatives of $P$ with respect to the canonical kinetic term $X$ one finds
\be
\frac{\lambda}{\Sigma} = \frac{1 + \frac{2}{3}X\frac{P_{,XXX}}{P_{,XX}}}{2 + \frac{P_{,X}}{X P_{,XX}}},
\ee 
so that in order for $|{\lambda}/{\Sigma}| \gg 1$ we require
\be
\left| X^2 P_{,XXX} \right|  \gg \left| P_{,X}  \right|.
\ee
Large non-Gaussianities sourced by $\lambda / \Sigma$ (or equivalently $f_X$) thus open up the doors for constraining $P_{,XXX}$.

Adapting the scheme introduced by~\cite{beanrecon} we can therefore construct a general form for any action that gives rise to a large non-Gaussian amplitude $\cA$ and is consistent with constraints on $X P_{,X}$, $X^2 P_{,XX}$ and $X^3 P_{,XXX}$. We find:
\bea \label{fxaction} 
\tilde{P}(X,\phi) &=& q(X,\phi)+P\left({X_0},\phi \right)-q\left({X_0},\phi \right)\nonumber
\\
&+&
\left[P_{,X} \left({X_0},\phi \right)-q_{,X} \left({X_0},\phi \right)\right] \left(X-{X_0} \right)\nonumber
\\
&+&{1\over2}\left[P_{,XX} \left({X_0},\phi \right)-q_{,XX}
\left({X_0},\phi \right)\right] \left(X-{X_0} \right)^2 \nonumber
\\
&+&{1\over6}\left[P_{,XXX} \left({X_0},\phi \right)-q_{,XXX}
\left({X_0},\phi \right)\right] \left(X-{X_0} \right)^3. 
\eea
(where $q$ is an arbitrary function of $\phi$ and $X$) so that higher derivatives along $X$ remain unconstrained. $X_0$ can be fixed by a gauge choice, associated with field redefinitions $\phi \to \tilde{\phi} = g(\phi)$ (\cite{beanrecon} use the gauge $X(N_e) = 1/2$ where $N_e$ is the number of e-folds of inflation). Choosing a value for $X$ at a specific time in this way is also equivalent to choosing a normalization of $\phi$. Having obtained the general action~\eqref{fxaction} one can thus write down a theory with any given value for $\lambda/\Sigma$ as discussed in section~\ref{shapes}, since $P_{,XXX}$ is not constrained in any other way than through the value taken by $\lambda/\Sigma$.

Figure~\ref{concreteshapes} shows that the generic shape generated by $\left|f_X \right| \gg 1$ is equilateral. In fact much more so than in the DBI case, since the $n_s$-dependent local contribution is confined to the extreme squeezed limit. As shown in detail in section~\ref{shapes} deviations from slow-roll strongly suppress the amplitude.

\subsection{Bimetric theories} \label{bimetric}

Bimetric theories~\cite{joao2,bimetricng,bekenstein} posit that gravity and matter are minimally coupled to two different metrics, related by a single dynamical scalar degree of freedom. Schematically they are governed by an action of the form:
\be
S=\frac{M_{Pl}^2}{2}\int
d^4 x {\sqrt {-g}}\, R[g_{\mu\nu}] + \int d^4 x {\sqrt {-{\hat
g}}}\, {\cal L}_m[{\hat g}_{\mu\nu},\Phi_{Matt}].
\ee
where $g_{\mu\nu}$ and $\hat{g}_{\mu\nu}$ are the gravity and matter metrics respectively. The two metrics may be seen as independent representations of the local Lorentz group, so that different Lorentz transformations must be used to transform measurements made with matter and gravity. For this reason no causality paradoxes arise even if the invariant speed of one metric appears superluminal with respect to the other metric~\cite{joao2,vikman}. This is in contrast to plain tachyonic matter~\cite{kinney2}\footnote{
A similar argument has been used for single metric, single field inflationary scenarios. If the field serves as a time dependent background establishing a preferred coordinate frame, no causality paradoxes come about because perturbations only travel superluminally in the preferred frame then~\cite{vikman}.}.

Disformal bimetric theories have been considered in the context of 
cosmological structure formation, following from:
\be
{\hat g}_{\mu\nu}=g_{\mu\nu}-B(\phi)\partial_\mu\phi\partial_\nu\phi\; .
\ee
It was shown that the minimal such theory (constant, positive $B$) results in scale-invariant curvature perturbations~\cite{joao2}, 
with a distinctive non-Gaussian signature~\cite{bimetricng}. Departures from scale-invariance and their associated non-Gaussian signatures were also examined in~\cite{bimetricng}. Projecting the scalar field action in the the Einstein (gravity) frame one obtains~\cite{joao2,bimetricng} an action of the DBI type
\be
S = \int d^4 x \sqrt{- g}  \left(\frac{R}{2} + \sqrt{1 + 2B(\phi)X}\frac{1}{B(\phi)} - \frac{1}{B(\phi)} + V(\phi)\right) \label{Sgrav}\; .
\ee
For a solution to the horizon problem, $B(\phi) = -f(\phi)>0$, so this is sometimes labelled ``anti-DBI'' (also see~\cite{Mukhanov:2005bu}). Obviously a positive $B$ cannot be generated from brane world considerations, due to the signature of space-time.  But the results derived in the previous sections can be easily applied to this class of theories.

Bimetric models produce a unique non-Gaussian signature. Taking the small tilt ($n_s - 1 \ll 1$) and large speed of sound $c_s \to \infty$ limits, we find~\cite{bimetricng}  
\begin{align}
{\cal A} \ = &\  \left(\frac{k_1 k_2 k_3}{2 K^3}\right)^{n_s -1} \left[- \frac{1}{8}\sum_i k_i^3 + \frac{1}{K} \sum_{i<j} k_i^2 k_j^2 -\frac{1}{2 K^2} \sum_{i\neq j}k_i^2 k_j^3\, \right. \nonumber \\
&\ + (n_s -1)\left(- \frac{1}{8}\sum_i k_i^3 - \frac{1}{8} \sum_{i\neq j} k_i k_j^2 + \frac{1}{8}k_1 k_2 k_3  + \frac{1}{2 K} \sum_{i<j} k_i^2 k_j^2- \frac{1}{2 K^2} \sum_{i\neq j}k_i^2 k_j^3\right)\nonumber \\ 
&\ \left. + {\cal O}\left(\frac{1}{c_s^2}\right) \right] \; . \label{ouramplitude}
\end{align}
The corresponding expression for $\fnl$ is given in~\ref{fnlbimetric}. Any dependence on $\epsilon$ therefore drops out and the amplitude is uniquely fixed by the  spectral index $n_s$, establishing a consistency relationship between $n_s$ and ${\cal A}$. This holds great promise for testing bimetric models with non-Gaussian constraints, in contrast to the inflationary relation between scalar spectral index $n_s$ and the tensor-to-scalar ratio $r$ in inflation. Non-Gaussian features in the scalar fluctuations are more likely to be observationally accessible in the near future than tensor modes.

Figure~\ref{concreteshapes} confirms that the amplitude $\cA$ is independent of $\ep$ for bimetric models. Note that the peaks in the squeezed limit are solely due to breaking of scale-invariance and thus are a direct measure of $n_s$. They may still be observationally elusive, if confined too far into the squeezed limit. We emphasize that no significant constraints arise for bimetric theories from $n_{NG}$, as the tilt of the three-point function is strongly suppressed for $c_s \gg 1$.

\section{Conclusions} \label{conc}

In this paper we have computed the amplitude $\cA$, size $\fnl$, shape and running $n_{NG}$ of the non-Gaussianity to be found in single field models, without assuming slow-roll  or exact scale-invariance of the scalar power spectrum. The main results are given in section~\ref{nongaussianity} and boil down to equations~\eqref{amplitudes1} and~\eqref{amplitudes2}, which are general expressions for the non-Gaussian amplitude in single field models. All other results, relating to  the size, shape and running of non-Gaussianities, follow from there. 
We have shown that observational constraints allow significant violations of slow-roll conditions $\ep \ll 1$ and $\eps \ll 1$ in inflationary single field models. As such we derived explicit bounds on slow-roll parameter $\ep$ in non-scale-invariant and non-slow-roll single field scenarios. We found models with $\ep$ as large as $\sim 0.3$ satisfying all present bounds.

A much richer non-Gaussian phenomenology can be associated with single field models which violate the slow-roll conditions. We established the following results:
\begin{itemize}
\item The size of non-Gaussianity $\fnl$ is generically reduced when departing from the slow-roll regime. This implies that models with a small speed of sound $c_s$, which violate observational constraints in the slow-roll limit, can still be allowed when considering fast-roll scenarios.
\item Wherever large non-Gaussianities do arise for single field models, they typically have an equilateral shape in the slow-roll limit. We showed that fast-roll models can give rise to shapes with larger orthogonal and local contributions (away from the extreme squeezed limit). For example, cases where the amplitude $\cA$ peaks in the folded limit $2k_1 = 2k_2 = k_3$ are constructed explicitly in section~\ref{shapes}. Except for large local non-Gaussianity near the squeezed limit, {\it which remains a smoking gun of multifield models}, all the interesting shapes can therefore be generated in single field models.
\item Fast-roll models generically give rise to a large running of non-Gaussianity $n_{NG}$ with scale. However, contrary to models with $n_s = 1$, these can be red-tilted. We predict exact values for the running in terms of the model parameters in section~\ref{running}.
\item For bimetric models $(\ep > 1)$ the consistency relation derived in~\cite{bimetricng} was confirmed: the amplitude $\cA$ is a fixed function of $n_s$. An equilateral shape was obtained with small local corrections in the squeezed limit, proportional to $(n_s - 1)$. We also derived expressions for $\fnl$, showing explicitly that $\fnl \sim {\cal O}(1)$ and positive, and we showed that $n_{NG} \sim 0$ for these models.
\item Three explicit model examples of theories that exhibit the phenomenology discussed above were constructed in section~\ref{concrete}.
\end{itemize}

In the future we hope to take this work further and investigate a number of issues. Having computed general single field expressions, it would be interesting to compute non-slow-roll non-Gaussian signatures for further specific models that fall within the single field remit. Furthermore, consistency relations between two- and three-point correlation functions exist in some limits~\cite{malda,consistency,rpconsistency,Huang:2010up,Cheung:2007sv,Ganc:2010ff,Gruzinov:2004jx,Sugiyama:2011jt,Suyama:2007bg}. In analogy with the bimetric scenario one should be able to find constraints uniquely relating observables such as $n_s$ to the full non-Gaussian amplitude {\it and} shape~\cite{consistencyshape}. This would be a useful extension of results reported in the literature regarding limits of the amplitude~\cite{malda,consistency,rpconsistency,Cheung:2007sv,Gruzinov:2004jx}. One could also extend the computations presented in this paper to cover slow-roll violations associated with $\eta \sim {\cal O}(1)$. This could help cure problems encountered when embedding cosmological models into e.g. supergravity. Finally it will be an interesting task to investigate higher order propagator corrections to the three-point function as discussed in section~\ref{fullamp}.

Non-Gaussianity is one of the best tools available for testing theories of primordial structure formation. With upcoming experiments promising to measure not just the size, but the full functional space mapped out by $\cA$, constraints from the shape of non-Gaussianity may become even more important in differentiating models. In this paper we have demonstrated that single scalar fields can source large non-Gaussianities with a much richer phenomenology than the negative $\fnl$, pure equilateral type non-Gaussianity produced by e.g. DBI inflation. Only time will tell, but while future observations might show that slow-roll single field models are ruled out, the simplest explanation could also very well still be a single field model. One that violates slow-roll.

\hspace{2cm}

{\textbf{{\begin{large}Acknowledgements:\end{large}}}} We would like to thank B. Hoare, N. Mekareeya, A. Mozaffari, D. Mulryne, D.Thomas, J. Yearsley and especially F. Piazza for helpful discussions, comments and correspondence. JN is supported by an STFC studentship. JM thanks STFC for support.

\appendix
\section{Appendix}
\subsection{Details on the computation of the three-point function} \label{details}

Here we give some further details showing how the amplitudes presented in section~\ref{fullamp} are computed explicitly. For further details we refer to~\cite{bimetricng}. 

\subsubsection{Mode functions $u_k(y)$}
In section~\ref{spectral} we found that the mode functions for scalar perturbations are given by
\begin{equation}
u_k(y) =  \frac{c_s^{1/2}}{a M_{\rm Pl} \sqrt{2 \epsilon}} v_k(y) =  \frac{c_s^{1/2}}{a M_{\rm Pl} 2^{3/2}} \sqrt{\frac{\pi}{\epsilon}} \sqrt{-y} H^{(1)}_\nu (- k y) .
\end{equation}
We now expand at $|k y| \ll 1$ to obtain an approximate expression for the Hankel function $H^{(1)}_\nu (- k y)$
\begin{equation} \label{Hankelexp}
H^{(1)}_\nu (- k y) = - i \, \frac{2^\nu \Gamma(\nu) (- k y)^{-\nu}}{\pi} [1+ i k y + {\cal O}(k y)^2] e^{-i k y}\,.
\end{equation}
which gives the following approximate expression for $u_k$,
\begin{equation} \label{umode}
u_k(y) \approx - i \, \frac{H (\epsilon + \es -1)}{2 M_{\rm Pl} \sqrt{ c_s k^3 \epsilon}} \left(\frac{- k y}{2}\right)^{3/2 - \nu} (1 + i k y) e^{- i k y} ,
\end{equation}
where we have used $\Gamma(\nu \approx 3/2) \approx \sqrt{\pi}/2$ as scale invariance constrains $\nu - 3/2 \ll 1$. One can check that modes freeze out in the $y\rightarrow 0$ limit by comparing with~\eqref{twotimes} and noting that
\begin{equation} \label{andamento}
\frac{H}{c_s^{1/2}} \sim (-y)^{\nu - 3/2}\, .
\end{equation}
Differentiating $u_k(y)$ with respect to $y$ one also finds
\begin{equation} \label{uprime}
u'_k(y) \approx -i \, \frac{H (\epsilon + \es -1)}{2 M_{\rm Pl} \sqrt{ c_s k^3 \epsilon}} \left(\frac{- k y}{2}\right)^{3/2 - \nu} k^2 y \ e^{- i k y}.
\end{equation}

\subsubsection{Field redefinitions and the cubic action}
For the final term in the cubic action~\eqref{action3}, $2 f(\zeta) \frac{\delta L}{\delta\zeta}|_1$, we have
\begin{eqnarray}
\left.\frac{\delta
L}{\delta\zeta}\right\vert_1 &=& a
\left( \frac{d\partial^2\chi}{dt}+H\partial^2\chi
-\epsilon\partial^2\zeta \right) ~,
\end{eqnarray}
\begin{eqnarray} \label{redefinition}
f(\zeta)&=&\frac{\eta}{4c_s^2}\zeta^2+\frac{1}{c_s^2H}\zeta\dot{\zeta}+
\frac{1}{4a^2H^2}[-(\partial\zeta)(\partial\zeta)+\partial^{-2}(\partial_i\partial_j(\partial_i\zeta\partial_j\zeta))] \nonumber \\
&+&
\frac{1}{2a^2H}[(\partial\zeta)(\partial\chi)-\partial^{-2}(\partial_i\partial_j(\partial_i\zeta\partial_j\chi))] ~,
\end{eqnarray}
where $\partial^{-2}$ is the inverse Laplacian. Since $\frac{\delta L}{\delta\zeta}|_1$ is proportional to
the linearized equations of motion, it can be absorbed by a field redefinition
\begin{eqnarray}
\zeta \rightarrow \zeta_n+f(\zeta_n) ~.
\end{eqnarray}

\subsubsection{The individual three-point amplitude computation}

Here we show the computation of ${\cal A}_{\zeta \dot\zeta^2}$ in detail. All the other amplitudes are computed along the same lines. The corresponding three-point correlator is
\begin{multline}
\langle \zeta(\textbf{k}_1)\zeta(\textbf{k}_2)\zeta(\textbf{k}_3)\rangle_{\zeta \dot\zeta^2}\, =\,
i (2 \pi)^3 \delta^3(\kk_1+\kk_2+\kk_3) u_{k_1}(y_{\rm end})u_{k_2}(y_{\rm end})u_{k_3}(y_{\rm end}) \\ \times
\int_{-\infty+i\varepsilon}^{y_{\rm end}} {\rm d} y \frac{c_s}{a} \frac{a^3 \epsilon}{c_s^4} (\epsilon - 3 + 3 c_s^2)
u_{k_1}^*(y) \frac{d u_{k_2}^*(y)}{d y}\frac{d u_{k_3}^*(y)}{d y} + {\rm perm.} + {\rm c.c.}
\end{multline}
The subscript ``end'' means that the quantity has to be evaluated at the end of the structure forming (scaling) phase. Making use of \eqref{twotimes}, \eqref{umode} and \eqref{uprime} we take some time-independent combinations outside the integral and evaluate them at $y = y_{\rm end}$:
\begin{multline} \label{threepointintegral}
\langle \zeta(\textbf{k}_1)\zeta(\textbf{k}_2)\zeta(\textbf{k}_3)\rangle_{\zeta \dot\zeta^2}\, =\, i (2 \pi)^3 \delta^3(\kk_1+\kk_2+\kk_3)
\frac{H_{\rm end}^6 \comb^6 2^{6\nu - 9}}{4^3 {M_{Pl}}^{4}\, \epsilon^{2}\, c_{s\, \rm end}^{\, 3}}\frac{(k_1 k_2 k_3)^{3 -2\nu}}{\Pi_j k_j^3} |y_{\rm end}|^{6\left(\frac{3}{2} - \nu\right)} \\
\times \int_{-\infty+i\varepsilon}^{{y_{\rm end}}} {\rm d}y (\epsilon - 3 + 3 c_s^2)
\frac{a^2}{{c}_s^{3}} (1 - i k_1 y) k_2^2 k_3^2 y^2 e^{i  K y}  + {\rm perm.} + {\rm c.c.}\,.
\end{multline}
Note that we have dropped a factor of $\Pi_j (1 + i k_j y_{end})e^{-i K y_{end}}$ as this will be negligibly small in the limit $|ky| \ll 1$ where the truncated Hankel function expansion \eqref{Hankelexp} is valid.
Evaluating the integral and re-expressing variables in terms of quantities calculated at sound horizon crossing (when, by convention, $y = K^{-1}$) we obtain
\begin{multline}
\langle \zeta(\textbf{k}_1)\zeta(\textbf{k}_2)\zeta(\textbf{k}_3)\rangle_{\zeta \dot\zeta^2}\, =\, (2 \pi)^3 \delta^3(\kk_1+\kk_2+\kk_3)
\frac{{\bar H}^4 (\epsilon + \es -1)^4 2^{6 \nu - 9}}{16 {M_{Pl}}^4 \epsilon^{2}\, {\bar c}_s^{4}}\frac{1}{\Pi_j k_j^3} \frac{(\Pi_j k_j^3)^{3-2\nu}}{K^{9 - 6 \nu}} \\
\times \frac{k_2^2 k_3^2}{K}
\left\{(\epsilon-3) \cos\frac{\alpha_2 \pi}{2} \Gamma(1+\alpha_2)\left[1+ (1+ \alpha_2) \frac{k_1}{K}\right] + 3 {\bar c}_s^2 \cos\frac{\alpha_1 \pi}{2} \Gamma(1+\alpha_1)\left[1+ (1+\alpha_1)\frac{k_1}{K}\right] \right\}+{\rm sym}. 
\end{multline}
where, as defined before in~\eqref{alphas},
\begin{eqnarray}
\alpha_1 = 3 - 2\nu = \frac{2 \epsilon + \es}{\es + \epsilon -1} \qquad ; \qquad \alpha_2  =  \frac{2 \epsilon - \es}{\es + \epsilon -1}.
\end{eqnarray}

\subsubsection{A useful integral}

Here we report a results from~\cite{kp} for evaluating integrals of the type~\eqref{threepointintegral}. These integrals are typically of the form
\begin{equation} \label{integral}
{\cal C} = \int_{-\infty + i\varepsilon}^{y_{\rm end}} {\rm d} y \left(\frac{y}{y_{\rm end}}\right)^\gamma (- i y)^n e^{i K y}\, .
\end{equation}
For $\gamma + n >-2$ the imaginary part of \eqref{integral} is convergent as $y_{\rm end}\rightarrow 0$. Note especially the $i\ep$ term which is necessary in order to regularize the integral. We can now approximately extend the upper limit of integration to $0$, which amounts to neglecting terms of higher order in $(k |y_{\rm end}|)$ consistent with the Hankel function expansion~\eqref{Hankelexp} we adopted earlier. One finds
\begin{equation} \label{Cconvergent}
{\rm Im}\, {\cal C} = - (K |y_{\rm end}|)^{-\gamma} \cos\frac{\gamma \pi}{2} \Gamma(1+ \gamma + n) K^{-n-1}\, .
\end{equation}

\subsection{General expressions for $\fnl$ and $n_{NG}$} \label{generalfnl}

Here we give a general expression for $\fnl$ as derived from the full amplitudes~\eqref{amplitudes1} and~\eqref{amplitudes2}.
We remember that $\fnl$ was defined as
\be
f_{\rm NL} = 30\frac{{\cal A}_{k_1=k_2=k_3}}{K^3}\,.
\ee
Substituting in the correct amplitude one obtains
\begin{eqnarray}
&& \fnl^{general} = \nonumber \\
&& \frac{5 \cdot 2^{-3-n_s} 3^{-2-3 n_s}}{\ct (1+\ep) (n_s - 2)} \Big(36 \ct (1+{\ep}) (-27+{n_s} (13+7 {n_s})) {\Gamma}[{n_s}] \text{sin}\left[\frac{{n_s} \pi }{2}\right]+ \nonumber\\
&& 16 \ct (1+{\ep})^2 {\Gamma}[2+{n_s}] \text{sin}\left[\frac{{n_s} \pi }{2}\right]+\left(-9 \left(9 {\ep}^3 ({n_s} - 2) ({n_s} - 1)-3 {\ep}^2 (38+5 ({n_s} - 7) {n_s})+  \right.\right. \nonumber \\
&& \left.\left.  8 {\ep} \left(21-22 {n_s}+8 {n_s}^2\right)-4 (33+{n_s} (5{n_s} - 31))\right) {\Gamma}\left[\frac{(3 {\ep} - 1) ({n_s}-2)}{1+{\ep}}\right]+ \right. \nonumber \\
&& \left.16 (1+{\ep})^2 ({f_X} - 1) {\Gamma}\left[\frac{4-4 {\ep}-{n_s}+3 {\ep} {n_s}}{1+{\ep}}\right]\right) \text{sin}\left[\frac{({\ep} (8-3 {n_s})+{n_s}) \pi }{2 (1+{\ep})}\right]\Big).
\end{eqnarray}
We also found that $\fnl$ can be expressed as
\be
\ln |f_{\rm NL}| = \ln |{\cal C}_1| + \ln  |1 - {\cal C}_2 \bar{c}_s^{-2}|,
\ee 
so that the non-Gaussian tilt $n_{NG}$ was given by
\be
n_{NG} - 1 = \frac{-2 \eps}{\eps + \ep - 1} \frac{{\cal C}_2}{\cbt - {\cal C}_2},
\ee
where ${{\cal C}_i} = {{\cal C}_i}(n_s,\ep,f_X)$. These functions are
\begin{eqnarray}
{\cal C}_1 (n_s,\ep,f_X) \sim \frac{5  (-243+{n_s} (121+67 {n_s}+4 {\epsilon} (1+{n_s}))) {\Gamma}[{n_s}] \text{sin}\left[\frac{{n_s} \pi }{2}\right]}{ 2^{1+{n_s}} 3^{2+3 {n_s}} (n_s - 2)},
\end{eqnarray}
\begin{align}\begin{split}
{\cal C}_2 (n_s,\ep,f_X) &\sim -\frac{5 \cdot 2^{-3-{n_s}} 3^{-2-3 {n_s}}}{(1+{\ep}) ({n_s} - 2)} \left(9 \left(9 {\epsilon}^3 ({n_s} - 2) ({n_s} - 1)-3 {\epsilon}^2 (38+5 ({n_s} - 7) {n_s})+ 
\right.\right.\\&\left.\left.\quad\quad{}
8 {\epsilon} \left(21-22 {n_s}+8 {n_s}^2\right)-4 (33+{n_s} (5 {n_s} - 31))\right) {\Gamma}\left[\frac{(3 {\epsilon} - 1) ({n_s} - 2)}{1+{\epsilon}}\right]- 
\right.\\&\left.\quad\quad{}
16 (1+{\epsilon})^2 ({f_X} - 1) {\Gamma}\left[\frac{4-4 {\epsilon}-{n_s}+3 {\epsilon} {n_s}}{1+{\epsilon}}\right]\right) \text{sin}\left[\frac{({\epsilon} (8-3 {n_s})+{n_s}) \pi }{2 (1+{\epsilon})}\right].
\end{split}\end{align}

\end{document}